\documentclass[12pt]{article}

\let\plural=\relax
\expandafter\ifx\csname shortcite\endcsname\relax\fi
\def\possnewtheorem#1#2{
\expandafter\ifx\csname #1\endcsname\relax
\newtheorem{#1}{#2}
\fi
}
\possnewtheorem{theorem}{Theorem}
\possnewtheorem{lemma}{Lemma}
\possnewtheorem{definition}{Definition}
\possnewtheorem{example}{Example}
\def\proof{\noindent {\sl Proof.\ \ }}
\def\qed{\hfill{\boxit{}}
  \ifdim\lastskip<\medskipamount \removelastskip\penalty55\medskip\fi}
\long\def\boxit#1{\vbox{\hrule\hbox{\vrule\kern3pt
                  \vbox{\kern3pt#1\kern3pt}\kern3pt\vrule}\hrule}}
\long\def\nop#1{}

\def\txt{txt}\ifx\outformat\txt\let\ttytty=\par\fi
\long\def\ttytex#1#2{#1}

\def\draft{\begingroup\em [draft]}
\def\enddraft{[/draft]\endgroup}
\newif\draft
\let\enddraft=\fi

\ifx\ttytty\par
\def\explain{\mbox{=>}}

\else
\def\explain{\Rightarrow}

\fi
\def\abduct{\mathrm{abduct}}
\def\focus{\mathrm{focus}}
\def\summarize{\mathrm{summarize}}

\ifx\citeyear\Undefined\let\citeyear=\cite\fi

\def\nojournal{\iftrue\begingroup\par\em=============[nojournal]=============\par}
\def\endnojournal{\par=============[/nojournal]=============\par\endgroup\fi}
\def\nojournal{\iffalse}
\let\endnojournal=\fi

\def\l{\langle}
\def\r{\rangle}
\def\true{{\sf true}}
\def\false{{\sf false}}
\def\P#1{\mbox{$\Pi^p_{#1}$}}
\def\S#1{\mbox{$\Sigma^p_{#1}$}}

\title{Abductive forgetting}
\author{Paolo Liberatore%
\footnote{Sapienza University of Rome, Via Ariosto 25, 00185, Rome, Italy
email:\tt liberato@diag.uniroma1.it}}

\begin{document}

\maketitle

\draft

{\bf focus on R = concentrate on R, like blowing up a part}

include related explanations

\

{\bf summarize on R = give an overview on a topic}

omit explanations not on the subject

\enddraft

\draft
\newpage

\section{sinossi}

\begin{enumerate}

\item {\bf introduction.tex}

\begin{itemize}

\item nel forget classico si mantengono le conseguenze

ma e' possibile voler mantenere altro

esempi di cosa altro si puo' mantenere: mutua consistenza, comportamento della
revisione (analogo a strong persistence in asp), spiegazioni abduttive

\item le ipotesi da dimenticare vengono rimosse dalle spiegazioni; altrimenti
si verificano casi come ``a, b e c spiegano d'' che diventano ``d e'
inspiegabile''

\item invece sulle conseguenze cancellare quelle da dimenticare dalle
spiegazioni o cancellare del tutto le spiegazioni che le contengono sono due
soluzioni che hanno senso in contesti diversi; esempi: diagnosi mediche e
materiale didattico

\item a seconda dei casi, forget puo' non essere possibile se deve produrre una
formula proposizionale

\item riassunto dell'articolo

\end{itemize}

\item {\bf definitions.tex}

definizione di spiegazione abduttiva e di spiegazioni minime

assunzioni: manifestazioni e ipotesi non intersecanti, manifestazioni non
vuote, contenimento stretto implica minore stretto

si spiega ancora perche' la definizione di forget e' esistenziale sulle ipotesi

le due definizioni: mantenere le spiegazioni con manifestazioni da non
dimenticare, tutte o alcune; perche' entrambe hanno senso, ognuna in contesti
diversi

entrambe le definizioni producono un insieme di spigazioni E=>M, che poi puo' o
meno essere rappresentabile come un altro problema di abduzione

\item {\bf difference.tex}

differenze fra forget abduttivo e consequenziale

\begin{itemize}

\item producono risultati diversi

\item forget consequenziale non distinguie fra focus e summarizing

\end{itemize}

\item {\bf expression.tex}

definizione di espressibilita' di forget con una formula proposizinale e
controesempi

\begin{itemize}

\item definizione di espressibilita' del forget come abduzione con una formula
proposizionale

\item forget puo' non esprimersi come abduzione da una formula proposizionale

\item se si esprime per focus allora si esprime anche per summarize e viceversa

\end{itemize}

\item {\bf algorithm.tex}

sistema per generare una formula G(S) a partire da un insieme di spigazioni
S={E=>M}

si dimostra che se esiste una formula che ha esattamente le spiegazioni in S,
allora S sono anche le spiegazioni di G(S)

qui si parla di tutte le spiegazioni

\item {\bf condition.tex}

un insieme di spiegazioni S={E=>M} e' esattamente l'insieme delle spiegazioni
di una qualche formula proposizionale se e solo se valgono le seguenti due
condizioni:

\begin{enumerate}

\item congiunzione: se E spiega M1 ed M2, allora spiega anche M1M2

\item overreaching monotony: se E spiega m, allora anche un suo sovrainsieme
spiega m, se esiste un sovrainsieme di entrambi che spiega qualcosa; i
contenimenti possono non essere stretti

\end{enumerate}

si dimostrano non essere sempre vere quando S e' l'insieme delle spiegazioni
del forget

\item {\bf complexity.tex}

complessita' del problema di esprimere forget con una formula proposizionale

il problema corrisponde alla validita' delle due condizioni per le spiegazioni
dopo il forget; in effetti poi si guarda il problema della non espressibilita'
perche' e' piu' chiaro da spiegare

\begin{itemize}

\item il contrario di ciascuna delle condizioni e' in sigma3, quindi anche il
problema

\item la hardness viene fatta vedere su (and); la dimostrazione viene poi
trasformata nella hardness della non espressibilita' mostrando che se la QBF e'
falsa allora forget si esprime

\item poi la hardness viene anche fatta vedere per (monot) quando si
considerano solo manifestazioni singole; che poi in efftti anche questa e' una
dimostrazione di hardness del problema della non esprimibilita', dato che e'
come assumere che (and) sia rispettato

\end{itemize}

\item {\bf minimal.tex}

spiegazioni minime, usando un ordinamento <= quasi arbitrario, con l'unico
vincolo che sottoinsieme stretto implica minore stretto

si definisce G(S) per questo caso

due condizioni (less) e (min) insieme sono necessarie e sufficenti alla non
esprimibilita'; in questo sono analoghe a (and) e (monot), ma molto diverse

\begin{enumerate}

\item less: l'insieme non contiene due spiegazioni delle stesse manifestazioni
di cui una e' minore dell'altra

\item min: questa condizione esprime in termini soltanto dell'insieme di
spiegazioni minime certe condizioni che forzano un insieme di ipotesi a essere
una spiegazione mimima di un insieme di manifestazioni

\end{enumerate}

si dimostra che un insieme di spiegazioni minime e' supportato da qualche
formula se e solo se non soddisfa nessuna delle due condizioni

complessita': solo membership, la non espressibilita' e' in sigma4

\item {\bf variables.tex}

il risultato del forget puo' non essere esprimibile come l'insieme delle
spiegazioni di una formula proposizionale

questo puo' succedere perche' una spiegazione ab=>m diventa a=>m, che non e'
esattamente la stessa cosa; la prima indica che ab spiegano m, la seconda che
a, in certe condizioni, spiega b; queste condizioni possono contraddirsi fra
loro e contraddire le ipotesi; il primo caso produce una violazione di (and),
il secondo a una violazione di (monot)

una soluzione e' quella di specificare queste interazioni

uno dei modi di farlo e' introducendo della variabili da non contare

tecnicamente, si esprime il forget di X come una formula da cui dimenticare Y

chiaramente si puo' sempre fare usando come Y=X, il punto e' che le condizioni
aggiuntive su X vengono semplificate; invece delle clausole di partenza, che
riflettono la conoscenza attuale, si specifica soltanto il minimo che serve per
realizzare le interazioni fra le "certe condizioni" e fra queste e le ipotesi

un esempio che viene considerato e' quello di introduzione di una sola
variabile aggiuntiva; indica che una interazione e' possibile, ma deve essere
molto semplice

\begin{itemize}

\item viene mostrato un esempio che fa vedere una formula da cui dimenticare n
variabili non si esprime in logica proposizionale ma si esprime con una singola
nuova variabile; quindi non e' la stessa cosa che reintrodurre la variabile
dimenticata

\item esiste una differenza qualitativa fra una variabile e piu' variabili: con
una variabile riesco a esprimere certi insiemi che violano (and), ma nessuno
che viola (monot); per quelli servono almeno due variabili

\item viene dato un algoritmo che dovrebbe trovare la formula che esprime
forget con una sola variabile aggiuntiva se esiste; dovrebbe funzionare, ma e'
senza dimostrazione

\end{itemize}

\item {\bf default.tex}

una alternativa a introdurre nuove variabili e' quella di usare una logica piu'
potente di quella proposizionale

in questa sezione si considera default logic: un insieme di ipotesi E spiega
delle manifestazioni se queste sono vere in almeno una estensione della teoria
di default con E aggiunto; e' la semantica credolous perche' con quella
skeptical non si ottiene nessun vantaggio rispetto alla logica proposizionale

esiste in effetti un modo molto semplice per esprimere qualsiasi insieme di
spiegazioni in questo modo; pero' richiede un insieme di regole di default che
uguale al numero di spiegazioni dell'insieme! funziona solo grazie alla
separazione di ipotesi e manifestazioni

esiste un modo migliore, piu' intuitivo; ma richiede default non normali, e di
usare una semantica specifica per i default (rational?)

\item {\bf intersecting.tex}

ipotesi e manifestazioni sono sempre assunte disgiunte;
cosa cambia quando non lo sono

in questo caso, e' possibile avere spiegazioni a catena come a=>b e b=>c

per tutte le spiegazioni e forse per quelle minime per sottoinsieme non cambia
niente, perche' in entrambi i casi devono valere a->b e b->c e quindi anche
a->c; ma nel caso della minimalita' per cardinalita' questo discorso non vale
piu', nel senso che anche se a e' una spiegazione minima di b e c lo e' di c,
non e' detto che a sia una spiegazione minima di c

e infatti la formula G(S) non funziona piu' quando si considerano spiegazioni
minime

viene proposto un sistema simile a G(S), ma con dei tentativi; uno di questi
potrebbe essere la formula che esprime forget se ne esiste una

\item {\bf related.tex}

uso di forget per trovare le spiegazioni, non per ridurre il linguaggio

inferenza cumulativa definita da abduzione ha un concetto simile a quello
di una formula che supporta delle spiegazioni date

sistema generale di forget da stato doxastico, che puo' codificare anche
il forget abduttivo

trovare le variabili da dimenticare come fatto nel forget consequenziale
si applica anche a quello abduttivo

\item {\bf conclusions.tex}

discussione dei risultati:

\begin{itemize}

\item definizioni multiple sono comuni nel forget in altre logiche

\item cosa indica la l'assenza di una formula che supporta il forget
abduttivo: che le variabili da dimenticare sono molto correlate con quelle
da mantenere, e che dimenticare complica la cose invece di
semplificarle

\item problemi aperti:

\begin{itemize}

\item miglioramento dell'efficienza dell'algoritmo, che in questo articolo
serve solo nelle dimostrazioni teoriche

\item grandezza della formula che support il forget

\item altre forme di default logic che supportano forget sempre

\item lo stesso per altre estensioni della logica proposizionale

\item ricerca di spiegazioni senza rappresentare forget con una formula,
partendo invece dai dati di partenza: la formula iniziale e le variabili
da dimenticare

\end{itemize}

\end{itemize}

\item {\bf other.tex}

spiegazioni aggiuntive possibili

estensioni possibili

dettagli sui lavori correlati

\end{enumerate}

\newpage

\enddraft

\begin{abstract}

Abductive forgetting is removing variables from a logical formula while
maintaining its abductive explanations. It is carried in two alternative ways
depending on its intended application. Both differ from the usual forgetting,
which maintains consequences rather than explanations. Differently from that,
abductive forgetting from a propositional formula may not be expressed by any
propositional formula. A necessary and sufficient condition tells when it is.
Checking it is \P{3}-complete%
\nojournal
, and is in \P{4} if minimality of
explanations is required
\endnojournal
. A way to guarantee expressibility of abductive
forgetting is to switch from propositional to default logic.
Another is to introduce new variables.

\end{abstract}

\section{Introduction}
\label{section-introduction}

Logical forgetting is restricting a logical formula on a subset of its
constituent elements, such as its
variables~\cite{bool-54,delg-17,eite-kern-19}. It has been
extensively studied in many settings~\cite{%
lin-01,lang-etal-03,moin-07,
lin-reit-94,
boto-etal-17,
leit-17,gonc-etal-21,
kone-etal-09,
wang-etal-15,
baum-bert-22}. 
It is also known as variable elimination in the context of automated
reasoning~\cite{davi-putn-60,dech-rish-94,subb-prad-04} and by its dual concept
of uniform interpolation in modal and description logic~\cite{kone-etal-09}.

The common form of logical forgetting is to remove variables from a
propositional formula while maintaining its consequences on the remaining ones.
The consequences of `A and B are true' are 'A is true' and `B is true'. Only
the first survives forgetting `B'.

The three key ingredients of logical forgetting come from its very definition.
Propositional forgetting is ``given a {\em propositional formula (Point 1)},
reduce {\em its variables (Point 2)} while maintaining {\em its consequences
(Point 3)} on the remaining ones.'' The three points are the three key
ingredients. Changing them gives different versions of forgetting.

\begin{itemize}

\item rather than a propositional formula, forget from a first-order
formula~\cite{lin-reit-94}, a modal logic formula~\cite{kone-etal-09}, from
circumscription~\cite{wang-etal-15}, from an ontology~\cite{boto-etal-17}, a
logic program~\cite{leit-17,gonc-etal-21}, an argumentation
framework~\cite{baum-etal-20}.

\item rather than the variables, restrict the literals~\cite{lang-etal-03}, the
objects~\cite{delg-17}, the subformulae~\cite{fang-etal-18};

\item rather than the consequences, maintain something else.

\end{itemize}

The last point lacks examples because it is the least explored direction of
research. An exception is strong persistence in Answer Set Programming (ASP):
maintaining the consequences is not enough, maintaining the semantics of the
formula after adding new parts is also required. This is one ``something else''
to maintain: the consequences after additions to the formula.

Inference and additions are not the only logical operations.
Checking what is possible is another. What explains something is another. What
changes after removals or additions is another. Each operation gives a version
of forgetting.

\begin{description}

\item[what is possible:]

reduce the alphabet of a propositional formula while
maintaining the formulae that are consistent with it;

\item[what explains something:]

reduce the alphabet of a propositional formula while
maintaining its abductive explanations~\cite{eite-gott-95-a};

\item[what holds after removal or revision:]

reduce the alphabet of a propositional formula while maintaining its
consequences after belief a contraction or revision~\cite{rodr-etal-11}.

\end{description}

This article is about the second extension: maintain what the formula explains.

Abduction is finding explanations of manifestations. A formula expresses the
relations between the possible manifestations (variables standing for
observable conditions) and the hypotheses that could explain them (other
variables that imply them).

\begin{example}

Medical diagnosis is finding the conditions that explain the symptoms.
For example, a condition A may explain D and E. Abduction frames this by a
formula that contains or otherwise entails a clause
{} ``A implies D and E''.
Its abuctive explanations of D includes A because adding A to the formula makes
it entail D while maintaining consistency.

\end{example}

Forgetting is restricting the context in some way, usually by removing
variables, to simplify reasoning or to focus attemption on a specific part of
the context. The two cases do not differ in propositional logic, but they do in
abduction; alternative definitions of forget are common in other
logics~\cite{gonc-etal-21,eite-kern-19}.

\begin{example}

A doctor is diagnosing a patient. The currently available test results only
tell that D is the case. Since A explains D and E, a test for E is prescribed,
but not yet done. Temporarily neglecting E does not rule out A as an
explanation. ``A explains D'' remains.

\end{example}

A different application of forgetting leads to a different outcome.

\begin{example}

A professor is writing some teaching material on the same medical topic. The
first chapter is about the basics, which include A and D, but not E. A following
chapter will cover E. To avoid the confusion of first exposing ``A explains D''
and then correcting it into ``A explains D and E'', this diagnosis is delayed
to the later chapter altogheter.

\end{example}

Both notions differ from classical forgetting in propositional logics, which
maintains the consequences of a formula instead of the explanations.

\begin{example}

The formula ``A,B and C imply D'' does not entail ``A and B imply D''.
Therefore, it is not maintained by classically forgetting C, which maintains
consequences.

This is not the case for explanations. A patient is not checked for condition C
because of a lack of time, but A and B cannot be ruled out as the causes of D.
A professor has not yet explained C, but needs to cover A, B and D, including
that A and B are possible reason for D.

\end{example}

In both cases, A and B are still possible reasons for D, even if D is not a
consequence of A and B alone. Maintaining the consequences removes it.
Maintaining the explanations keeps it. The contrary would be that a symptom D
that is explanained by A, B and C becomes unexplanaible when C is to be
neglected. The lack of time to perform a test on C would leave a doctor with no
explanation and no cure. The need to delay the exposition of C would force the
professor to overlook an important connection between A, B and C.

Maintaining the consequences is called consequential forgetting to distinguish
it from abductive forgetting, which maintains the explanations. In turn,
forgetting the hypotheses that could not be checked is called focusing;
forgetting to clarify writing is called summarizing. The doctor focuses on what
can be checked. The professor summarizes the topic.

Focusing and summarizing are the two forms of abductive forgetting. They
maintain some explanations and remove parts of others.

Forgetting produces explanations. Abduction expresses explanations in logic.
The original explanations comes from abduction from a propositional logic. The
resulting ones may not, they may not come from any propositional formula.

This is common in logics other than propositional logic. For example,
forgetting a predicate in first-order logic may not be first-order
definable~\cite{lin-reit-94}. Forgetting symbols from the description logics
ALC may not give an ALC formula~\cite{wang-etal-09}. Strongly persistent
forgetting from a logic program may be impossible~\cite{gonc-etal-16-a}. Some
modal logics do not have uniform interpolants; therefore, they do not always
allow forgetting~\cite{fang-etal-19}. Abductive forgetting is like them:
forgetting from a propositional formula may not give any propositional formula.

%
%
%
%

Forgetting something from a formula in a certain language may not be
representable within the same language. Whether it does for a specific formula
is the expressibility problem: checking whether abductive
forgetting from a specific propositional formula gives the set of explanations
of some propositional formula.

An algorithm computes a formula that expresses abductive
forgetting if one exists. A necessary and sufficient condition is provided, and
the complexity of the problem is established.

Abductive forgetting may not be expressed in propositional logic, but it is in
the more expressive default logic~\cite{reit-80,besn-89,anto-99}. Alternatively, introducing new variables solves the problem within propositional logic.

The article is organized as follows.
{}Section~\ref{section-definitions}
fixes the formalism and gives the two definitions of forgetting;
{}Section~\ref{section-difference}
proves they differ from the usual form of forgetting that maintains the
consequences instead of the abductive explanations;
{}Section~\ref{section-expression}
defines the problem of expressing abductive forgetting;
{}Section~\ref{section-algorithm}
presents an algorithm for forgetting when possible;
{}Section~\ref{section-condition}
shows a necessary and sufficient condition for the representability of
forgetting within propositional logic;
{}Section~\ref{section-complexity}
is about the complexity of the problem;
{}\nojournal%
{}Section~\ref{section-minimal}
extends the analysis to minimal explanations;
{}\endnojournal%
{}Section~\ref{section-default}
shows that default rules always allow forgetting;
{}
introducing new variables also does, as shown in
{}Section~\ref{section-variables}%
{}\nojournal%
{}; Section~\ref{section-intersecting}
tells what changes when hypotheses and effects are not disjoint%
{}\endnojournal%
.
Connections with related work are described in
{}Section~\ref{section-related},
discussion of the results in
{}Section~\ref{section-conclusions}.

\section{The two forms of abductive forgetting}
\label{section-definitions}

\draft

\begin{itemize}

\item definition of abduction, etc.

\item motivate again and formalize the two definitions of forgetting: focusing
and summarizing

examples of the difference

the existential choice on the hypotheses is motivated again, with examples;
when using each is told again, with examples

\end{itemize}

\enddraft

Abduction works on a propositional formula built over an alphabet that includes
two subsets: the hypotheses and the manifestations, assumed
disjoint~\cite{eite-gott-95-a,mehe-bate-06}%
{}\nojournal~%
; how this assumption affects the given result is discussed in
Section~\ref{section-intersecting}
{}\endnojournal.

The propositional formula represents all available knowledge about the
relationships between hypotheses and manifestations. It is written as a set of
clauses like $\{ab \rightarrow c, b \rightarrow d\}$, which is the same as
$(\neg a \vee \neg b \vee c) \wedge (\neg b \vee d)$. The head of a clause $ab
\rightarrow c$ is $c$, its body $ab$. This notation is only used for clauses,
sets of literals implying a literal; it is not used for implications between
formulae like $A \wedge B \rightarrow C$ if $A$, $B$ and $C$ are formulae.

A subset of hypotheses $E$ may or may not explain a nonempty subset of
manifestations $M$. It does if its union with the formula is consistent and
entails them. Such an explanation is written $E \explain M$. It is supported by
a formula $F$ if $F \cup E$ is consistent and entails $M$. This is denoted $F
\models E \explain M$.

Which explanations are supported depends not only on $F$ but also on the set of
all hypotheses and manifestations. The hypotheses $E$ of a specific explanation
$E \explain M$ are some of the possible hypotheses $I$; in the same way, the
manifestations $M$ are some of the possible manifestations $C$.

\begin{definition}
\label{explanation}

An {\em explanation} over hypotheses $I$ and manifestations $C$ is a pair of
propositional sets of variables $E \subseteq I$ and $M \subseteq C$ with $M
\not= \emptyset$, written $E \explain M$.

\end{definition}

The set of all possible hypotheses $I$ and manifestations $C$ are two sets of
variables. They are assumed disjoint unless stated otherwise: $I \cap C =
\emptyset$.

Explanations are $E \explain M$, while implications are $a \rightarrow b$.
Different symbols denote different concepts.

Formally, abduction is finding explanations out of a propositional formula
given the sets of all hypotheses and manifestations.

\begin{definition}
\label{abduction-frame}

An {\em abduction frame} is a triple $\l F,I,C \r$ where $F$ is a propositional
formula and $I$ and $C$ are two disjoint sets of variables respectively
denoting the set of all possible hypotheses and manifestations.

\end{definition}

The formal definition of support of an explanation follows.

\begin{definition}
\label{support}

An abduction frame $\l F,I,C \r$ {\em supports} an explanation $E \explain M$
over the set of all possible hypotheses $I$ and manifestations $C$ if $F \cup
E$ is consistent and entails $M$. This condition is written $\l F,I,C \r
\models E \explain M$. The set of explanations supported by an abduction frame
$\l F,I,C \r$ is denoted $\abduct(\l F,I,C \r)$.

\end{definition}

The set of all hypotheses $I$ and all manifestations $C$ are implicit in this
article: they are given for the formula before forgetting; they are $I \cap R$
and $C \cap R$ after forgetting, where $R$ is the set of variables to remember.
This makes $I$ and $C$ fixed in both cases. They are omitted for simplicity:
$\l F,I,C \r$ supports an explanation over $I$ and $C$ is shortened to $F$
supports an explanation. The sets $I$ and $C$ are implicit.

The notation $F \models E \explain M$ for support cannot be confused with $F
\models A$ for propositional entailment because $E \explain M$ is an
explanation while $A$ is a propositional formula.

To simplify notation, the explanation $\{a,b,c\} \explain \{d,e\}$ is written
$abc \explain de$.

\nojournal

The set of explanations supported by a formula may include both $a \explain c$
and $ab \explain c$. While the second is still a valid explanation, it may be
considered less relevant as it includes the unnecessary hypothesis $b$.
Similarly, $a \explain e$ may be preferred to $bcd \explain e$ because of its
smaller number of hypotheses: one instead of three.

Two definitions of minimality are considered: by set containment or by
cardinality~\cite{paul-93,eite-gott-95-a}. They only select the supported
explanations $E \explain M$ such that $E' \explain M$ is not supported by any
$E'$ less than $E$, where ``less'' is either $E' \subset E$ or $E'$ comprising
fewer hypotheses than $E$. An arbitrary order $<$ can be used instead of these
two. This article assumes that $E \subset E'$ implies $E < E'$. This is the
case for the two specific orderings above.

\endnojournal

Summary of the assumptions:

\begin{itemize}

\item $M$ is not empty: $E \explain \emptyset$ is not an explanation;

\item hypotheses and manifestations are disjoint: $E \cap I = \emptyset$%
{}\nojournal%
what happens when lifting this assumption is shown in
Section~\ref{section-intersecting}%
{}\endnojournal%
.


\end{itemize}

\

The notation $abc \explain de$ tell that $a$, $b$, and $c$ explain $d$ and $e$
according to the formula. The available knowledge says that if $d$ and $e$ are
the case, a possible reason is that $a$, $b$ and $c$ are as well.

What if the hypothesis $c$ and the manifestation $e$ are not of interest?

As outlined in the introduction, forgetting a hypothesis like $c$ and
forgetting a manifestation like $e$ differ. The first has a unique treatment;
the second does not.

\begin{description}

\item[Forgetting a hypothesis.]

\

The explanation $abc \explain de$ turns into $ab \explain de$ when $c$ is not
of interest for whichever reason. For example, reading ``not of interest'' as
``a detail to neglect'', neglecting that $c$ is a part of the cause of $d$ and
$e$ means that it is left out. Reading ``not of interest'' as ``an information
that is not of any use anyway'', $c$ is removed from the explanation as
useless. Regardless of the use of forgetting, $abc \explain de$ turns into $ab
\explain de$ when forgetting $c$.

The alternative is to remove $abc \explain de$ altogether as an explanation.
This would leave $d$ unexplainable if $abc$ is its only explanation. Even if
$d$ has another cause $fgh$, removing $abc \explain de$ hides $ab$ as an
alternative.

The patient is cured for $fhg$, even if it is dangerous in presence of $a$. The
student may only memorize $fhg$ as an explanation of $de$.

\item[Forgetting a manifestation.]

\

A tempting solution is to follow the same principle, turning $abc \explain de$
into $abc \explain d$ when forgetting $e$. This way of forgetting has some
ground. A medical manual may say that $a$, $b$, and $c$ cause $d$ and $e$.
Since $e$ cannot be established for now, it is better left out. Still, $a$, $b$
and $c$ explain $d$.

\begin{eqnarray*}
&& ab \explain de		\\
&& f \explain d			\\
&& fgh \explain de
\end{eqnarray*}

Since $e$ is unknown, it may be true or false. If it is false, $f$ is the
smallest explanation of $d$. If it is true, $f$ does not explain it alone,
without $g$ and $h$; the alternative explanation $ab$ is smaller. Since $e$ is
not know, none of the two alternatives can be excluded.

The doctor diagnosing $d$ and $f$ later teaches students about the illness $d$.
If it occurs alone, it is the smallest possible explanation of $f$. Only the
uncommon complication $e$ makes $a$ and $b$ a smaller alternative. For a quick
overview, the alternative is left out of that lecture.

\end{description}

Abductive forgetting is either focusing (like enlarging a detail of a
photograph with a magnifying glass) and summarizing (like outlining a topic).
Focusing removes forgotten manifestations. Summarizing removes explanations.

\begin{definition} \label{forget-intersect}

Focusing an abduction frame $\l F,I,C \r$ on a set of variables $R$ gives the
following set of explanations.

\begin{eqnarray*}
\lefteqn{\focus(\l F,I,C \r,R) =} \\
&&
	\{E \explain M \mid
		E \subseteq I \cap R ,~
		M \subseteq C \cap R ,~
		M \not= \emptyset ,~ \\
&&
		\exists E' \subseteq I \backslash R ,~
		\exists M' \subseteq C \backslash R ,~
		EE' \explain MM' \in \abduct(\l F,I,C \r)
	\}
\end{eqnarray*}

\end{definition}

\begin{definition}
\label{forget-contains}

Summarizing an abduction frame $\l F,I,C \r$ on a set of variables $R$ gives
the following set of explanations.

\begin{eqnarray*}
\lefteqn{\summarize(\l F,I,C \r,R) =} \\
&&
	\{E \explain M \mid
		E \subseteq I \cap R ,~
		M \subseteq C \cap R ,~
		\exists E' \subseteq I \backslash R ,~
		EE' \explain M \in \abduct(\l F,I,C \r)
	\}
\end{eqnarray*}

\end{definition}

Defining focusing and summarizing on the variables to remember $R$ instead of
those to forget simplifies the technical treatment, but the concept is the
same: focusing or summarizing on some variables is forgetting the others.

The hypotheses $I$ and manifestations $C$ in the abductive frame are fixed.
Consequently, they are also fixed after forgetting: $I \cap R$ and $C \cap R$.
They are left implicit, simplifying notation to $\focus(F,R)$ and
$\summarize(F,R)$.

Focusing and summarizing produce explanations and not propositional formulae
like consequentially forgetting $x$ from a propositional formula $F$ produces
the propositional formula $F[\true/x] \vee F[\false/x]$. Such a propositional
formula may not exists for focusing and summarizing, as proved by
Theorem~\ref{unsupported}.

Neither focusing nor summarizing make $E \explain \emptyset$ an explanation. It
is a consequence in the second definition and is explicitly stated in the
first. Without this constraint, forgetting $m$ would turn $ab \explain m$ into
$ab \explain \emptyset$. An explanation of something to be completely forgotten
is remembered. Still worse, it is remembered as an explanation of nothing: $ab
\explain \emptyset$ means that $a$ and $b$ explain why nothing is observed.

Both focusing and summarizing remove hypotheses from explanations. Focusing
also removes manifestations, summarizing removes whole explanations. In the
other way around, focusing simplifies each explanation $E \explain M$ to $E
\cap R \explain M \cap R$; summarizing simplifies $E \explain M$ to $E \cap R
\explain M$ if $M \subseteq R$, and removes it altogether otherwise. The latter
removes the explanations that the former shortens.

\begin{theorem}
\label{theorem-containment}

The containment $\summarize(F,R) \subseteq \focus(F,R)$ holds for every formula
$F$ and set of variables $R$.

\end{theorem}

\proof An explanation $E \explain M$ is in $\summarize(F,R)$ if some other
explanation $EE' \explain M$ is in $\abduct(F)$ with $E' \subseteq I \backslash
R$. A consequence of $EE' \explain M \in \abduct(F)$ is $M \not= \emptyset$
since explanations of no manifestations are excluded; another is $EE' \explain
MM' \in \abduct(F)$ with $M' = \emptyset$. Since $M'$ is empty, it is trivially
contained in $C \backslash R$. All requisites for $E \explain M$ being in
$\focus(F,R)$ are met.~\qed

Both focusing and summarizing are forms of forgetting. When their difference is
not important, such as when they coincide because all manifestations are
remembered, the generic term forgetting is used in their place.

\begin{theorem}
\label{coincide}

If $C \subseteq R$ then $\focus(F,R) = \summarize(F,R)$.

\end{theorem}

\proof Focusing $\focus(F,R)$ is defined as follows.

\[
	\{E \explain M \mid
		E \subseteq I \cap R ,~
		M \subseteq C \cap R ,~
		M \not= \emptyset ,~
		\exists E' \subseteq I \backslash R ,~
		\exists M' \subseteq C \backslash R ,~
		F \models EE' \explain MM'
	\}
\]

Since $C$ is contained in $R$, none of its elements is outside $R$. In terms of
sets, $C \backslash R$ is empty. The only subset $M'$ of an empty set is the
empty set: $M' = \emptyset$. The definition of $\focus(F,R)$ can therefore be
rewritten as:

\[
	\{E \explain M \mid
		E \subseteq I \cap R ,~
		M \subseteq C \cap R ,~
		M \not= \emptyset ,~
		\exists E' \subseteq I \backslash R ,~
		F \models EE' \explain M
	\}
\]

The definition of $F \models EE' \explain M$ includes $M \not= \emptyset$,
which can be therefore be removed from the definition of the set.

\[
	\{E \explain M \mid
		E \subseteq I \cap R ,~
		M \subseteq C \cap R ,~
		\exists E' \subseteq I \backslash R ,~
		F \models EE' \explain M
	\}
\]

This is the definition of $\summarize(F,R)$.~\qed

\section{Abductive and consequential forgetting}
\label{section-difference}

\draft

differences between abductive and consequential forgetting

- they may produce different results

- consequential forget does not distinguish between focusing and summarizing

\enddraft

Abductive forgetting differs from consequential forgetting. An example is
{} $F = \{ab \rightarrow x\}$,
where the hypotheses are
{} $I = \{a, b\}$,
the manifestations
{} $C = \{x\}$
and
{} $b$ is forgotten.
Both focusing and summarizing produce $a \explain x$%
{}\nojournal%
, which is also minimal
{}\endnojournal%
. This explanation is supported by the formula $\{a \rightarrow x\}$.
Consequential forgetting gives an empty formula instead. Replacing $b$ with
false turns $F = \{ab \rightarrow x\}$ into $\true$. As a result, the
disjunction $F[\true/b] \vee F[\false/b]$ is $\true$ as well.

Consequential forgetting can be defined as Boole's conditionalization:
forgetting $x$ from $F$ is $F[\true/x] \vee F[\false/x]$~\cite{lang-etal-03}.
This definition being syntactical is only a minor difference with abductive
forgetting: consequential forgetting is the same as the set of models over the
variables to remember that satisfy $F$ when expanded with arbitrary values of
the variables to forget~\cite{lang-etal-03}. The major difference is that they
are exactly the models of a propositional formula: $F[\true/x] \vee
F[\false/x]$. This is not the case with abductive forgetting: the resulting
explanations may not be produced by abduction from any propositional formulae.

Consequential forgetting does not distinguish between looking at a detail
(focusing) and giving an overview of a particular (summarizing). Both are given
as applications of forgetting with no distinction. For example, Eiter and
Kern-Isberner~\citeyear{eite-kern-19} wrote: ``not all information can be kept
and treated in the same way. [...] forgetting [...] helps us to deal with
information overload and to put a focus of attention''. This is an example of
summarizing, as a mean to omit details that are not important.
Botoeva~et~al.~\citeyear{boto-etal-17} wrote ``As an example, consider Snomed
CT, which contains a vocabulary for a multitude of domains related to health
case, including clinical findings, symptoms, diagnoses, procedures, body
structures, organisms, pharmaceuticals, and devices. In a concrete application
such as storing electronic patient records, only a small part of this
vocabulary is going to be used''. This is an example of focusing, as a mean to
limit information to what is necessary to a specific task.
Such applications of forgetting are commonly seen in the literature without any
distinction made. It is unnecessary since it would make no difference.

\section{Supporting forgetting}
\label{section-expression}

\draft

definition of forgetting being supported

forgetting may not be supported by a formula

does not depend on whether forget is focus or summarize

\enddraft

In both its forms, focusing and summarizing, forgetting produces a set of
explanations. This set may coincide with the set of explanations supported by
some formula. If it does, the formula is a representation of forgetting.

\begin{definition}
\label{forget-support}

Focusing $\l F,I,C \r$ on a set of variables $R$ is supported by the formula
$G$ if and only if
{} $\focus(\l F,I,C \r,R) = \abduct(\l G,I \cap R,C \cap R\r)$.
The same for summarizing.

\end{definition}

Contrarily to consequential forgetting, abductive forgetting may not be
supported by any formula.

\begin{theorem}
\label{unsupported}

Abductive forgetting may not be supported by any formula.

\end{theorem}

\proof The proof exhibits a counterexample formula. Abductive forgetting
produces a set of explanations that is not the set of explanations supported by
any formula.

The formula is
{} $F = \{ab \rightarrow x, ac \rightarrow y, bc \rightarrow \bot\}$,
the hypotheses
{} $I = \{a,b,c\}$,
the manifestations
{} $C = \{x,y\}$,
the variables to
{} forget $b$ and $c$,
those to
{} remember $R = \{a,x,y\}$.
The supported explanations are $ab \explain x$ and $ac \explain y$. Instead,
$abc \explain xy$ is not supported since the conjunction of $b$ and $c$ is
inconsistent with $F$.


Focusing and summarizing coincide since $C \subseteq R$ by
Theorem~\ref{coincide}. They turn $ab \explain x$ and $ac \explain y$ into $a
\explain x$ and $a \explain y$. They do not produce $a \explain xy$. Every
formula supporting the first two explanations also supports the third. The
conclusion is that no formula supports expactly the explanations $\{a \explain
x, a \explain y\}$.~\qed

This theorem applies to both focusing and summarizing: they may or may not be
supported by any formula. If they do, they do both. If they do not, neither
does.

\begin{theorem}
\label{focus-summarize-support}

Focusing is supported by a formula if and only if summarizing is.

\end{theorem}

\proof Focusing and summarizing change or remove explanations. They both remove
the variables to forget from all explanations, but they remove an explanation
only if its manifestations are respectively all or in part to be forgotten.
They differ only on explanations that contain some manifestations to be
remembered and some to be forgotten:
{} $EE' \explain MM'$
{} when forgetting $E' \cup M'$ and 
{} neither $M$ nor $M'$ is empty.
Focusing turns it into $E \explain M$, summarizing removes it. The original
$EE' \explain MM'$ is supported by the formula $F$. Therefore, $F \cup E \cup
E'$ is consistent and entails $M \cup M'$. As a result, it entails $M$ alone.
The same formula $F$ also supports $EE' \explain M$. Since $M$ is not empty and
comprises only variables to remember, both forms of forgetting turn it into $E
\explain M$~\qed

\section{Algorithmic generation of forgetting}
\label{section-algorithm}

\draft

\begin{itemize}

\item formula G(S) from a set of explanations S={E=>M}

\item Lemma~\ref{contained-body}: a formula where heads and bodies are disjoint
entails A->b if and only if it contains A'->b with A'c=A

\item Theorem~\ref{gs-for-all}: a formula supports exactly the explanations
S={E=>M} if and only if G(S) does

\end{itemize}

\enddraft

The explanations produced by a focusing or summarizing may or may not be the
explanations supported by a formula. An algorithm builds it if any exists.
Precisely, it builds a formula supporting exactly a given set of explanations,
including the ones produced by focusing and summarizing. The next section uses
the algorithm for proving a necessary and sufficient condition to the existence
of such a formula. The complexity results rely on this condition.

The algorithm synthesizes a formula that supports a given set of explanations
$S = \{E \explain M\}$ if one such formula exists. In its current form it is
meant to be used in proofs. It is not intended to be used in practice since it
is exhaustive on the explanations in $S$.

\begin{definition}
\label{supporting}

The tentative-supporting formula of a set of explanations $S$ over disjoint
hypotheses $I$ and manifestations $C$ is the formula $G(S)$ that comprises the
following clauses, where $m$ is a single manifestation:

\begin{enumerate}

\item a clause $E \rightarrow m$ for each $E \explain m \in S$;

\item a clause $E \rightarrow \bot$ for each $E \explain m \not\in S$ such that
$E' \explain m \in S$ for some $E' \subset E$.

\end{enumerate}

\end{definition}

No optimization is attempted, such as neglecting an explanation $E \explain m$
of $S$ when $E' \explain m$ is also in $S$ with $E' \subset E$. This would
produce a smaller formula, but complicates the proofs where the algorithm is
employed.

\begin{example}
\label{example-no-tentative-supporting}

The tentative-supporting formula of
{} $S = \{
{} 	a  \explain x,
{} 	ab \explain x,
{} 	 b \explain y
{} \}$
is
{} $G(S) = \{
{}	a  \rightarrow x,
{}	b  \rightarrow y,
{}	ab \rightarrow \bot
{} \}$.
Its first two clauses come from $a \explain x$ and $b \explain y$. Its third
clause comes from $b \explain y$ and the absence of $ab \explain y$. Its
supported explanations are $a \explain x$ and $b \explain y$. It does not
support $ab \explain x$ although this explanation is in $S$. No formula
supports $S$ because the only way $y$ could be explained by $b$ but not by $ab$
is because $ab$ is inconsistent, which would prevent $ab$ from explaining $x$.

\end{example}

Two similar sets of explanations are instead supported by their
tentative-supporting formula. The first adds $ab \explain y$ to the
explanations of Example~\ref{example-no-tentative-supporting}.

\begin{example}

The tentative-supporting formula of
{} $S = \{
{} 	a  \explain x,
{} 	ab \explain x,
{} 	 b \explain y,
{} 	ab \explain y
{} \}$
is $G(S) = \{a \rightarrow x, b \rightarrow y\}$. It does not contain $ab
\rightarrow \bot$ because $ab \explain y$ is in $S$. The explanations it
supports are exactly $S$.

\end{example}

The second example removes $ab \explain x$ from the explanations of
Example~\ref{example-no-tentative-supporting}.

\begin{example}

The tentative-supporting formula of
{} $S = \{
{} 	a \explain x,
{} 	b \explain y
{} \}$
is $G(S) = \{a \rightarrow x, b \rightarrow y, ab \rightarrow \bot\}$. Its
supported explanations are again $S$. The difference with the first set is the
absence of $ab \explain x$, which was part of why no formula supported the
first set.

\end{example}

The pivotal result about the tentative-supporting formula is that it is closed
on entailment: it entails a clause if and only if it contains that clause or
one entailing it.

\begin{lemma}
\label{contained-body}

If $I \cap C = \emptyset$ and $F$ only contains clauses $E \rightarrow m$ and
$E \rightarrow \bot$ with $E \subseteq I$ and $m \in C$, then $F$ entails a
clause if and only if it contains a clause with the same head if any and a
subset of its body.

\end{lemma}

\proof If $F$ contains a clause $E' \rightarrow m$ with $E' \subseteq E$, it
entails $E \rightarrow m$ because this clause is a superset of $E' \rightarrow
m$. The same for $E' \rightarrow \bot$.

In the other direction, if $F \models E \rightarrow m$ then $F \cup E \models
m$. This is the same as $h$ being generated by propagation from $F \cup E$. Let
$E' \rightarrow m$ be the last clause used in this derivation. By assumption,
$E'$ is a subset of $I$. Since no clause of $F$ contains any variable of $I$ in
the head, no variable in $E'$ follows from propagation. Therefore, they are all
in $E$. A similar argument proves the claim for the clauses $E \rightarrow
\bot$.~\qed


This lemma relies on the assumption that hypotheses and manifestations are
disjoint.
{}\nojournal
The effects of lifting this constraint are considered in
Section~\ref{section-intersecting}.
{}\endnojournal%
Also crucial is that no head is in a body, but this is guaranteed by the
definition of the tentative-supporting formula $G(S)$.

The main result about the tentative-supporting formula is that it supports the
set of explanations if and only if such a formula exists.

\begin{theorem}
\label{gs-for-all}

A formula $F$ supports exactly the explanations $S$ over disjoint hypotheses
and manifestations if and only if $G(S)$ does.

\end{theorem}

\proof The assumption is that every $E \explain M \in S$ satisfies $E \subseteq
I$ and $M \subseteq C$. The claim is that $\abduct(\l G(S),I,C \r) = S$ holds
if and only if a formula $F$ such that $\abduct(\l F,I,C \r) = S$ exists, where
every $E \explain M \in S$ satisfies $E \subseteq I$ and $M \subseteq C$.

If $G(S)$ supports $S$ then a formula supporting $S$ exists because it
is $G(S)$ itself.

The rest of the proof shows the converse: if a formula $F$ supports $S$, then
$G(S)$ does as well. The assumption is that $F$ supports $S$.

\

A first preliminary result is that $S$ contains an explanation $E \explain M$
if and only if it contains $E \explain m$ for every $m \in M$.

Since $F$ supports $S$, this set contains $E \explain M$ if and only if $F \cup
E$ is consistent and entails $M$. This is the same as $F \cup E$ being
consistent and entailing every $m \in M$. By definition, it is also the same as
$F$ supporting $E \explain m$ for every $m \in M$. Thanks to the assumption
that $F$ supports $S$, this is also the same as $S$ containing $E \explain m$
for every $m \in M$.

\

A second preliminary result is that $F$ entails $G(S)$.

This is proved by showing that $F$ entails every clause of $G(S)$. A clause $E
\rightarrow m$ is in $G(S)$ if and only if $S$ contains $E \explain m$. Since
$F$ supports $S$, it supports its member $E \explain m$. The definition of
support includes $F \models E \rightarrow m$.

A clause $E \rightarrow \bot$ is in $G(S)$ if $E \explain m$ is not in $S$, but
$E' \explain m$ is for some $E' \subset E$. Since $F$ supports exactly $S$, it
supports $E' \explain m$ but not $E \explain m$. As result, $F \cup E'$ is
consistent and entails $m$, and either $F \cup E$ is inconsistent or does not
entail $m$. The second possibility, $F \cup E \not\models m$, is excluded since
$F \cup E$ is a superset of $F \cup E'$, which entails $m$. As a result, the
first possibility is the case: $F \cup E$ is inconsistent. This is the same as
$F \models E \rightarrow \bot$.

%
%

\

The claim that $G(S)$ supports $S$ is now proved by contradiction. The converse
is that either $G(S)$ does not support an explanation of $S$ or that it
supports an explanation that is not in $S$. Both cases are proved
contradictory.

\begin{itemize}

\item $G(S)$ does not support $E \explain M \in S$;

By the first preliminary result, $E \explain M \in S$ is the same as $E
\explain m \in S$ for every $m \in M$. By construction, $E \rightarrow m$ is in
$G(S)$ for every $m \in M$. A consequence is that $G(S)$ entails $E \rightarrow
M$.

By assumption, $G(S)$ does not support $E \explain M$. Yet, it entails $E
\rightarrow M$. Therefore, $G(S) \cup E$ is inconsistent. This is the same as
$G(S) \models E \rightarrow \bot$. Since $F$ entails $G(S)$ by the second
preliminary result, it also entails $E \rightarrow \bot$. As a result, $F \cup
E$ is inconsistent. Therefore, $F$ does not support $E \explain M$,
contradicting the assumption that $E \explain M$ is in $S$.

%
%

\item $G(S)$ supports an explanation $E \explain M \not\in S$;

The definition of support is that $G(S) \cup E$ is consistent and entails $E
\rightarrow M$. The latter is the same as $G(S) \models E \rightarrow m$ for
every $m \in M$. By Lemma~\ref{contained-body}, $G(S)$ contains a clause $E'
\rightarrow m$ with $E' \subseteq E$ for every $m \in M$. By the definition of
$G(S)$, this is possible only if $S$ contains an explanation $E' \explain m$
for every $m \in M$.

Two possibilities are explored: either $E \explain m$ is in $S$ for all $m \in
M$, or it is not for some. In the second case, $S$ contains $E' \explain m$ but
not $E \explain m$ with $E' \subseteq E$. The case $E'=E$ is not possible since
otherwise $E \explain m$ and $E' \explain m$ would be the same, while the
latter is in $S$ and the former is not. Therefore, $E'$ is strictly contained
in $E$. By construction, $G(S)$ contains $E \rightarrow \bot$. This contradicts
the assumption that $G(S)$ supports $E \explain M$. The conclusion is that $E
\explain m$ is in $S$ for every $m \in M$. By the first preliminary result
above, $S$ contains $E \explain M$, contrary to assumption.

%
%

\end{itemize}

The assumption that the explanations of $G(S)$ are not $S$ leads to
contradiction. The consequence is that $G(S)$ supports $S$.~\qed

The tentative-supporting formula may be exponentially larger than other
formulae supporting the same explanations. This is not a problem because its
motivation is theoretical: it is employed in some of the following proofs.

\section{Necessary and sufficient conditions}
\label{section-condition}

\draft

\begin{itemize}

\item definition of two conditions: the conjunctive condition and the
overreaching monotony condition

\item Theorem~\ref{all-iff}:
a set of explanations meets both conditions if and only if some formula
supports it

\end{itemize}

\enddraft

A necessary and sufficient condition is proved for a set of explanations being
supported by a formula. As in the previous section, the set of explanations may
be the result of forgetting, either focusing or summarizing, but not
necessarily. If it is, the condition specializes on whether forgetting is
expressed by a formula.

The first part of the condition is that manifestations are jointly explained if
and only if each is explained.

\begin{definition}[Conjunctive condition]
\label{and}

A set of explanations $S=\{E \explain M\}$ satisfies the conjunctive condition
if it contains both $E \explain M_1$ and $E \explain M_2$ if and only if it
contains $E \explain M_1 M_2$.

\end{definition}

If $S$ is the set of explanations supported by a formula, it satisfies this
condition.

\begin{lemma}
\label{satisfies-and}

The set of explanations supported by an arbitrary formula satisfies the
conjunctive condition.

\end{lemma}

\proof The definition of a formula $F$ supporting an explanation $E \explain
M_1M_2$ is that $F \cup E$ is consistent and entails $M_1 M_2$. The latter
is the same as $F \cup E$ entailing both $M_1$ and $M_2$. These conditions are
also the same as $F$ supporting both $E \explain M_1$ and $E \explain M_2$
since $F \cup E$ is consistent.~\qed

Abductive forgetting may not satisfy the conjunctive condition. A
counterexample is
{} $F=\{ab \rightarrow m, ac \rightarrow m', \neg b \vee \neg c\}$
where the hypotheses are
{} $I = \{a, b, c\}$
and the manifestations
{} $C = \{m, m'\}$.
The explanations supported by $F$ are $ab \explain m$ and $ac \explain m'$,
while $abc \explain mm'$ is not since $b$ and $c$ are not together consistent
with $F$. These three explanations do not violate the conjunctive condition
since they have different preconditions. More generally, the explanations
supported by a formula always satisfy the condition. This is not the case when
forgetting variables. Forgetting $b$ and $c$ turns the first two explanations
into $a \explain m$ and $a \explain m'$, while still not producing $a \explain
mm'$. Their preconditions are the same. The conjunctive condition is violated.

A set of explanations $S$ satisfying the conjunctive condition is completely
defined by its explanations with a single manifestation. The others are derived
by the rule that $E \explain M$ is in $S$ if and only if $E \explain m$ is in
$m$ for all $m \in M$.

While the conjunctive condition is necessary to $S$ being supported by a formula,
it is not sufficient. An additional condition is required.

\begin{definition}[Overreaching monotony condition]
\label{monotony}

A set of explanations $S = \{E \explain M\}$ satisfies {\em overreaching
monotony} if $E \explain m \in S$ and $E'' \explain m' \in S$ imply $E'
\explain m \in S$ when $E \subseteq E' \subseteq E''$.

\end{definition}

This is a sort of ``converging monotony'' or ``bilateral monotony'': $E
\explain m$ implies $E' \rightarrow m$, while $E'' \explain m'$ implies the
consistency of $E'$ with the formula. A consequence is $E' \explain m$. This is
always the case for the abductive explanations of a formula.

\begin{lemma}
\label{satisfies-monotony}

The set of explanations supported by an arbitrary formula satisfies the
overreaching monotony condition.

\end{lemma}

%

\proof The assumption is that a formula $F$ supports a set of explanations $S$.
The claim is that $S$ satisfies overreaching monotony: $E \explain m \in S$,
$E'' \explain m' \in S$ and $E \subseteq E' \subseteq E''$ imply $E' \explain m
\in S$.

%
%
%

A consequence of $E \explain m \in S$ is that $E \explain m$ is supported by
$F$. By definition, $F \cup E$ entails $m$. Since $F \cup E'$ is a superset of
$F \cup E$, it entails $m$ as well.

For the same reason, $E'' \explain m' \in S$ implies that $E'' \explain m'$ is
supported by $F$. This implies the consistency of $F \cup E''$. Since $F \cup
E'$ is a subset of $F \cup E''$, it is consistent as well.

The conclusions are that $F \cup E'$ is consistent and entails $m$. This
defines $F$ supporting $E' \explain m$. Since $S$ is the set of explanations of
$F$, it contains $E' \explain m$.~\qed

Overreaching monotony is defined on single manifestations only because it only
matters when the conjunctive property is satisfied.


Overreaching monotony is always met by the explanations supported by a formula,
but not always by forgetting. A counterexample is
{} $F = \{ab \rightarrow m, \neg b \vee \neg c, ac \rightarrow m'\}$
where the hypotheses are
{} $I = \{a, b, c, d\}$
and the manifestations
{} $C = \{m, m'\}$.
Its explanations are $ab \explain m$ and $ac \explain m'$. They do not violate
overreaching monotony because their preconditions are not contained one in the
other and their only superset $abc$ does not explain anything. Forgetting $b$
turns $ab \explain m$ into $a \explain m$ and leaves $ac \explain m'$
unaffected. Since $F$ supports neither $ac \explain m$ nor $abc \explain m$,
forgetting does not produce $ac \explain m$. This is a violation since $a
\subseteq ac \subseteq ac$.

The converse of the two lemmas is the case: the two conditions are not only
necessary for a set of explanations to be supported by some formula, they are
also sufficient.

\begin{theorem}
\label{all-iff}

A set of explanations over disjoint hypotheses and manifestations is supported
by a formula if and only if it satisfies the conjunctive and overreaching
monotony conditions.

\end{theorem}

\proof Lemma~\ref{satisfies-and} and~\ref{satisfies-monotony} tell that the set
of explanations of a formula satisfies both conditions.


The claim is proved by showing that if a set of explanations is not the set of
explanations supported by any formula, it violates either condition. In
particular, if it satifies the conjunctive condition then it violates the
overreaching monotony condition.

%

If $S$ is not the set of explanations of any formula, it is not the set of
explanations of $G(S)$ since this is a formula. Two cases are possible: either
$G(S)$ supports an explanation that is not in $S$, or it does not support an
explanation in $S$.



\begin{itemize}

\item $G(S)$ supports $E \explain m \not\in S$;




The definition of $G(S)$ supporting $E \explain m$ is that $G(S) \cup E$ is
consistent and entails $m$. The latter is the same as $G(S) \models E
\rightarrow m$. By Lemma~\ref{contained-body}, $G(S)$ contains a clause $E'
\rightarrow m$ with $E' \subseteq E$.

The containment of $E'$ in $E$ may be strict or not. The latter case $E' = E$
implies $E \rightarrow m \in G(S)$, which by construction implies $E \explain m
\in S$, which is not the case by assumption. The conclusion is that $G(S)$
contains $E' \rightarrow m$ with $E' \subset E$. By construction, $S$ contains
$E' \explain m$.

Since $S$ contains $E' \explain m$ and does not contain $E \explain m$ where
$E' \subset E$, by construction $G(S)$ contains $E \rightarrow \bot$. This
clause contradicts the consistency of $G(S) \cup E$.

\item $G(S)$ does not support $E \explain m \in S$;

%
%
%

Since $E \explain m$ is in $S$, by construction $G(S)$ contains $E \rightarrow
m$. Since $G(S)$ does not support $E \explain m$, either $G(S) \cup E$ is
inconsistent, or it does not entail $m$. The second case is not possible
because $G(S)$ contains $E \rightarrow m$. As a result, $G(S) \cup E$ is
inconsistent.

This is the same as $G(S) \models E \rightarrow \bot$. By
Lemma~\ref{contained-body}, $G(S)$ contains a clause $E' \rightarrow \bot$ with
$E' \subseteq E$. This clause is in $G(S)$ only if $S$ does not contain $E'
\explain m'$, but it contains $E'' \explain m'$ for some $E'' \subseteq E'$ and
some manifestation $m'$. At the same time, $S$ contains $E \explain m$ by
assumption. Also $E' \subseteq E$ is the case.

In summary, $S$ contains
{} $E'' \explain m'$ and
{} $E \explain m$,
it does not contain
{} $E' \explain m'$
and 
{} $E'' \subseteq E' \subseteq E$
holds. Overreaching monotony implies $E' \explain m' \in S$.

\end{itemize}~\qed

The conditions equate the existence of a formula supporting a given set of
explanations. If this set is the result of forgetting variables from a formula,
it is the existence of a formula supporting forgetting.

\section{Complexity}
\label{section-complexity}

\draft

\begin{itemize}

\item complexity of establishing whether some formula supports exactly the
explanations of forgetting

\item the necessary and sufficient condition is used in for both membership and
hardness

\item each of the two parts of the condition is in P3; the whole problem has
the same complexity

\item the conjunctive condition is P3-hard; the same lemma proves that the
whole problem has the same complexity

\item overreaching monotony is P3-hard; the same lemma proves in an alternative
way that the whole problem has the same complexity

\end{itemize}

\enddraft

Establishing whether abductively forgetting variables from a propositional
formula is supported by a formula is \P{3}-complete. While it is harder than
the building blocks of abduction, propositional satisfiability and entailment,
it is still within the polynomial hierarchy. It cannot probably be solved in a
polynomial amount of time, but requires only a polynomial amount of memory.
Furthermore, its complexity is similar to that of other problems in
abduction~\cite{eite-gott-95-a}.

The actual proof is made of two parts: membership to \P{3} and \P{3}-hardness.
Both rely on the necessary and sufficient condition proved in the previous
section. The second also proves the conjunctive condition \P{3}-hard. A
separate proof shows the same for overreaching monotony. Checking each
condition alone is \P{3}-hard.


\subsection{Membership}

An arbitrary set of explanations $S = \{E \explain M\}$ is supported by a
formula if and only if it meets both the conjunctive and overreaching monotony
conditions. If this set of explanations is the result of forgetting, this is
the problem of whether forgetting is expressed by a propositional formula.

A specific explanation $E \explain M$ results from forgetting if and only if
$E' \subseteq I \cap R$ and $M' \subseteq C \cap R$ exists such that $EE'
\explain MM'$ is supported. Depending on the definition of forgetting, $M \not=
\emptyset$ or $M' = \emptyset$ is also required.

Whether forgetting is focusing or summarizing does not matter, as
Theorem~\ref{focus-summarize-support} proves that the existence of a formula
supporting their result is the same
{}\nojournal%
when selecting all explanations and not only the minimal ones
{}\endnojournal%
. The following proofs are for the second definition because it is slightly
simpler, as it does not require a set $M'$ at all. An explanation $E \explain
M$ is in the result of forgetting if $ED \explain M$ where $D$ is an arbitrary
subset of $I \cap R$.


\begin{lemma}
\label{membership-and}

Checking whether $\summarize(F,R)$ satisfies the conjunctive property is in
\P{3}.

\end{lemma}

\proof The set of explanations $\summarize(F,R)$ violates the conjunctive
condition if either of the two following conditions is the case:

\begin{eqnarray*}
\exists E,M_1,M_2 .
&& E \explain M_1 \not\in \summarize(F,R)		\\
&& E \explain M_1M_2 \in \summarize(F,R)		\\
\exists E,M_1,M_2 .
&& E \explain M_1 \in \summarize(F,R)			\\
&& E \explain M_2 \in \summarize(F,R)			\\
&& E \explain M_1M_2 \not\in \summarize(F,R)
\end{eqnarray*}

The definition of $E \explain M \in \summarize(F,R)$ is
{} $\exists D \subseteq I \backslash R. ED \explain M \in \abduct(F)$.

The first of the two conditions is never the case. Its second part $E \explain
M_1M_2 \in \summarize(F,R)$ is the same as
{} $\exists D \subseteq I \backslash R . ED \explain M_1M_2 \in \abduct(F)$,
which is the same as the existence of a subset $D$ of $I \backslash R$ such
that
{} $F \cup E \cup D \not\models \bot$
and
{} $F \cup E \cup D \models M_1M_2$.
Since $F \cup E \cup D \models M_1M_2$ implies $F \cup E \cup D \models M_1$,
this implies
{} $\exists D .
{}  F \cup E \cup D \not\models \bot \mbox{ and }
{}  F \cup E \cup D \models M_1$,
which define $E \explain M_1 \in \summarize(F,R)$.

The second of the two conditions can be rewritten as follows.

\begin{eqnarray*}
\exists E,M_1,M_2 .
&& \exists D_1 \subseteq I \backslash R . ED_1 \explain M_1 \in \abduct(F) \\
&& \exists D_2 \subseteq I \backslash R . ED_2 \explain M_2 \in \abduct(F) \\
&& \forall D_3 \subseteq I \backslash R . ED_3 \explain M_1M_2 \not\in \abduct(F)
\end{eqnarray*}

Making $\abduct(F)$ explicit:

\begin{eqnarray*}
\exists E,M_1,M_2 .
&& \exists D_1 \subseteq I \backslash R .
	F \cup E \cup D_1 \not\models \bot \mbox{ and }
	F \cup E \cup D_1 \models M_1 \\
&& \exists D_2 \subseteq I \backslash R .
	F \cup E \cup D_2 \not\models \bot \mbox{ and }
	F \cup E \cup D_2 \models M_2 \\
&& \forall D_3 \subseteq I \backslash R .
	F \cup E \cup D_3 \models \bot \mbox{ or }
	F \cup E \cup D_3 \not\models M_1M_2
\end{eqnarray*}

The first two subconditions do not negate the third since the subset $D_1$,
$D_2$ and $D_3$ may differ from each other. An example is a first subcondition
satisfied only by a subset $D_1$ not entailing $M_2$, and a second subcondition
satisfied only by another subset $D_2$ not entailing $M_1$.

Checking entailment requires a universal quantifier, checking consistency
requires an existential one. The most alternations are in the third
subcondition:
{} $\exists E,\ldots \forall D_3 . F \cup E \cup D_3 \not\models M_1M_2$.
Three quantifiers, first existential: this proves membership in \S{3}.

This is the complexity class of checking whether the conjunctive condition is
violated. The class of checking whether it is met is its complement,
\P{3}.~\qed

Checking overreaching monotony has the same complexity: it is in \P{3}.

\begin{lemma}
\label{membership-monot}

Checking whether $\summarize(F,R)$ satisfies overreaching monotony is in \P{3}.

\end{lemma}

\proof Overreaching monotony is violated by $\summarize(F,E)$ if:

\begin{eqnarray*}
\exists E,E',E'',m,m'
&.& E \subseteq E' \subseteq E''		\\
&& E \explain m \in \summarize(F,R)		\\
&& E' \explain m \not\in \summarize(F,R)	\\
&& E'' \explain m' \in \summarize(F,R)
\end{eqnarray*}

The definition of $E \explain m \in \summarize(F,R)$ is
{} $\exists D \subseteq I \backslash R . ED \explain m \in \abduct(F)$.

\begin{eqnarray*}
\exists E,E',E'',m,m'
&.& E \subseteq E' \subseteq E''		\\
&& \exists D \subseteq I \backslash R . ED \explain m \in \abduct(F)		\\
&& \forall D \subseteq I \backslash R . E'D \explain m \not\in \abduct(F)	\\
&& \exists D \subseteq I \backslash R . E''D \explain m' \in \abduct(F)
\end{eqnarray*}

In turn,
{} $ED \explain m \in \abduct(F)$
is defined as
{} $F \cup E \cup D \not\models \bot$ and
{} $F \cup E \cup D \models m$.

\begin{eqnarray*}
\exists E,E',E'',m,m'
&.& E \subseteq E' \subseteq E''		\\
&& \exists D \subseteq I \backslash R .
	F \cup E \cup D \not\models \bot \mbox{ and }
	F \cup E \cup D \models m \\
&& \forall D \subseteq I \backslash R .
	F \cup E' \cup D \models \bot \mbox{ or }
	F \cup E' \cup D \not\models m \\
&& \exists D \subseteq I \backslash R .
	F \cup E'' \cup D \not\models \bot \mbox{ and }
	F \cup E'' \cup D \models m'
\end{eqnarray*}

Checking entailment requires a universal quantifier, checking consistency
requires an existential one. The most alternations are in the second part:
{} $\exists E,\ldots \forall D . F \cup E' \cup D \not\models m$.
Three quantifiers, first existential equals \S{3}.

This is the complexity of checking whether overreaching monotony is violated.
The checking whether it is met is its complement, \P{3}.~\qed

The two lemmas give an upper bound to the complexity of checking whether
forgetting is supported by some formula.

\begin{lemma}
\label{membership}

Checking whether $\summarize(F,R)$ is supported by a formula is in \P{3}.

\end{lemma}

\proof The problem amounts to checking the conjunctive and the overreaching
monotony conditions. Each is in \P{3} by Lemma~\ref{membership-and} and
Lemma~\ref{membership-monot}. Two checks of the same complexity have the same
complexity. The problem is therefore in \P{3}.~\qed

\subsection{Hardness}

The hardness of checking the existence of a formula supporting forgetting is
shown to be \P{3}-hard. The following lemma proves this claim by showing that
the conjunctive condition is \P{3}-hard.

\begin{lemma}
\label{hard-and}

The following two problems are \P{3}-hard:
checking whether $\summarize(F,R)$ satisfies the conjunctive property;
checking whether $\summarize(F,R)$ is supported by some formula.

\end{lemma}

\proof The hardness of the conjunctive condition is shown first. The reduction
is then shown to also prove the hardness of checking whether forgetting is
supported by some formula. Both results are proved by showing the reverse
problem \S{3}-hard.

The reduction for the conjunctive property is from the validity of a quantified
Boolean formula
{} $\exists X \forall Y \exists Z. F$.
Its variables are
{} $X = \{x_1,\ldots,x_n\}$,
{} $Y = \{y_1,\ldots,y_n\}$
and
{} $Z = \{z_1,\ldots,z_n\}$.
The corresponding formula $G$ is the following, where all variables not in $X
\cup Y \cup Z$ are new.

\begin{eqnarray*}
G &=&
XPS \cup XNS \cup XPN \cup XSS \cup XPX \cup XNX \cup YPY \cup YNY \cup \{FM\} \\
XPS &=& \{x_i^p x_i^s \rightarrow m_i \mid 1 \leq i \leq n\}		\\
XNS &=& \{x_i^f x_i^s \rightarrow m_i \mid 1 \leq i \leq n\}		\\
XPN &=& \{x_i^p x_i^f \rightarrow \bot \mid 1 \leq i \leq n\}		\\
XSS &=& \{x_i^s m_1 \ldots m_n \rightarrow \bot \mid 1 \leq i \leq n\} \\
XPX &=& \{x_i^p \rightarrow x_1 \mid 1 \leq i \leq n\}			\\
XNX &=& \{x_i^f \rightarrow \neg x_1 \mid 1 \leq i \leq n\}		\\
YPY &=& \{y_i^p \rightarrow y_1 \mid 1 \leq i \leq n\}			\\
YNY &=& \{y_i^f \rightarrow \neg y_1 \mid 1 \leq i \leq n\}		\\
FM &=&
F \vee
(\neg x_1^p \wedge \neg x_1^f) \vee \cdots \vee (\neg x_n^p \wedge \neg x_n^f) \vee
(\neg y_1^p \wedge \neg y_1^f) \vee \cdots \vee (\neg y_n^p \wedge \neg y_n^f)
\vee									\\
&& (m_1 \wedge \cdots \wedge m_n)
\end{eqnarray*}

The hypotheses $I$, manifestations $C$ and variables to remember $R$ are:

\begin{eqnarray*}
I &=& \{x_i^p, x_i^f, x_i^s, y_i^p, y_i^f \mid 1 \leq i \leq n\}	\\
C &=& \{m_i \mid 1 \leq i \leq n\}					\\
R &=& \{x_i^p, x_i^f, m_i \mid 1 \leq i \leq n\}
\end{eqnarray*}

The explanations $\summarize(G, R)${\plural} violate the conjunctive condition
when the quantified Boolean formula $\exists X \forall Y \exists Z . F$ is
true.

What do the parts of the formula do?

The first part $XPS \cup XNS \cup XPN$ comprises three clauses for each index
$i$:

\begin{eqnarray*}
&& x_i^p x_i^s \rightarrow m_i			\\
&& x_i^f x_i^s \rightarrow m_i			\\
&& x_i^p x_i^f \rightarrow \bot
\end{eqnarray*}

Each manifestation $m_i$ is explained by $\{x_i^p, x_i^s\}$ and by $\{x_i^f,
x_i^s\}$ but not by their union. When forgetting $x_i^s$, these explanations
turn into $x_i^p \explain m_i$ and $x_i^f \explain m_i$, while $\{x_i^p,
x_i^f\}$ and its supersets do not explain $m_i$.

This extends from single manifestations to sets of manifestations, with an
exception. For example, $\{m_1,m_2\}$ is explained by any combination of one
hypothesis among $x_1^p$ and $x_1^f$ and one among $x_2^p$ or $x_2^f$. The
exception is the set of all manifestations $\{m_1,\ldots,m_n\}$. The clauses of
$XSS$ rule out these explanations.

\[
x_i^s m_1 \ldots m_n \rightarrow \bot
\]

Regardless of how $m_1,\ldots,m_n$ are entailed, they contradict all variables
$x_i^s$. No explanation of $\{m_1,\ldots,m_n\}$ contains any $x_i^s$. The
arbitrary combinations of $x_1^p$ and $x_1^f$ no longer explain these
manifestations.

The conclusion is that all sets of manifestations are explained by an arbitrary
combination of $x_i^p$ and $x_i^f$, except the set of all manifestations
$\{m_1,\ldots,m_n\}$.

This is a violation of the conjunctive condition unless all such combinations of
$x_i^p$ and $x_i^f$ explain $\{m_1,\ldots,m_n\}$ through other clauses.

The other hypotheses to forget are $y_i^f$ and $y_i^f$. The clauses allowing
them to entail $m_1,\ldots,m_n$ are $XPX \cup XNX \cup YPY \cup YNY$:

\begin{eqnarray*}
&& x_i^p \rightarrow x_i			\\
&& x_i^f \rightarrow \neg x_i			\\
&& y_i^p \rightarrow y_i			\\
&& y_i^f \rightarrow \neg y_i			\\
\end{eqnarray*}

The only variables to remember among these are $x_i^p$ and $x_i^f$. A complete
combination of them, supplemented by a complete combination of $y_i^p$ and
$y_i^f$, forces a complete evaluation over the variables $X \cup Y$. Only such
evaluations falsify the central part of the subformula $FPN$:

\[
F \vee
(\neg x_1^p \wedge \neg x_1^f) \vee \cdots \vee (\neg x_n^p \wedge \neg x_n^f) \vee
(\neg y_1^p \wedge \neg y_1^f) \vee \cdots \vee (\neg y_n^p \wedge \neg y_n^f) \vee
(m_1 \wedge \cdots \wedge m_n)
\]

Only when the central part of this formula is false, and $F$ is falsified by
the evaluation over $X$ and $Y$ regardless of $Z$, the final part
{} $m_1 \wedge \cdots \wedge m_n$
is forced to be true. In the other way around, when an evaluation over $X$
exists such that for all values of $Y$ the formula $F$ is satisfiable,
{} $\{m_1,\ldots,m_n\}$
is not explained this way.

This proves that $\exists X \forall Y \exists Z . F$ equals
{} $\{m_1,\ldots,m_n\}$
not being explained by a combination of $x_i^p$ and $x_i^f$ for all $1 \leq i
\leq n$, in violation of the conjunctive condition.

The conclusion is that the conjunctive condition is falsified exactly when the
QBF is true. This proves that the conjunctive condition is \P{3}-hard.

%
%

The correspondence between the validity of the QBF and the violation of the
conjunctive condition can be rewritten as follows: if the conjunctive
condition is false, the QBF is true; if the QBF is true, the conjunctive
condition is false. The latter proves that forgetting is not supported by any
formula if the QBF is true. The only missing part is that overreaching monotony
is true when the QBF is false. Rather than proving that, a formula that
supports $\summarize(G,R)$ is shown when the QBF is false.

When the QBF is false, each $m_i$ is explained by $x_i^p$, $x_i^f$ and every
superset that does not contain two variables of the same index. These are the
explanations supported by the formula
{} $\{
{}	x_i^p \rightarrow m_i,
{}	x_i^f \rightarrow m_i,
{}	x_i^p x_i^f \rightarrow \bot
{}	\mid 1 \leq i \leq n
{} \}$.
This concludes the proof.~\qed

\subsection{Hardness of overreaching monotony}

The previous lemma shows that checking the conjunctive condition is \P{3}-hard.
This may suggest that this is the difficult part of the problem. This is not
the case: overreaching monotony is equally hard.

\begin{lemma}
\label{hard-monot}

Checking whether $\summarize(F,R)$ satisfies overreaching monotony is
\P{3}-hard.

\end{lemma}

\proof Reduction is from the validity of a quantified Boolean formula
{} $\exists X \forall Y \exists Z. F$
to the violation of overreaching monotony. The three sets of variables are
{} $X = \{x_1,\ldots,x_n\}$,
{} $Y = \{y_1,\ldots,y_n\}$
and
{} $Z = \{z_1,\ldots,z_n\}$.
The corresponding formula $G$ is the following, where all variables not in $X
\cup Y \cup Z$ are new.

\begin{eqnarray*}
G &=&
  \{AC\} \cup XPX \cup XNX \cup \{XC\} \cup YPY \cup YNY \cup
  \{FM\} \cup \{AB\} \\
AC &=& ac \rightarrow m							\\
XPX &=& \{x_i^p \rightarrow x_i \mid 1 \leq i \leq n\}			\\
XNX &=& \{x_i^n \rightarrow \neg x_i \mid 1 \leq i \leq n\}		\\
XC &=& a (x_1^p \vee x_1^n) \ldots (x_n^p \vee x_n^n) \rightarrow \neg c \\
YPY &=& \{y_1^p \rightarrow y_i \mid 1 \leq i \leq n\}			\\
YNY &=& \{y_1^n \rightarrow \neg y_i \mid 1 \leq i \leq n\}		\\
FM &=&
	F \vee \neg a \vee c \vee
	(\neg y_1^p \wedge \neg y_1^n) \vee \cdots \vee
	(\neg y_n^p \wedge \neg y_n^n) \vee
	m								\\
AB &=& ab \rightarrow m
\end{eqnarray*}

The hypotheses $I$, manifestations $C$ and variables to remember $R$ are:

\begin{eqnarray*}
I &=& \{a,b,c\} \cup \{x_i^p, x_i^n, y_i^p, y_i^n \mid 1 \leq i \leq n\} \\
C &=& \{m\}								\\
R &=& \{a,b\} \cup \{x_i^p, x_i^n \mid 1 \leq i \leq n\} \cup \{m\} 
\end{eqnarray*}

The claim is that $\exists X \forall Y \exists Z . F$ is true if and only if
overreaching monotony is violated by $\summarize(G,R)$. It is proved by linking
each evaluation over $X$ that satisfies $\forall Y \exists Z . F$ with a pair
of sets $E',E''$ that falsifies overreaching monotony for $m$, $m'$ and some
$E$.

Such a violation occurs when $\summarize(G,R)$ contains
{} $E \explain m$ and
{} $E'' \explain m'$, but not
{} $E' \explain m$
for some $E \subseteq E' \subseteq E''$. This is proved with $m'=m$, $E =
\{a\}$, $E'$ equal to $E$ with the addition of either $x_i^p$ or $x_i^n$
depending on the value of $x_i$ for each $i$ between $1$ and $n$, and $E''$
equal to $E'$ with the addition of $b$.

The explanation $E \explain m$ is in $\summarize(G,R)$ because $G$ contains $ac
\rightarrow m$ and $G \cup \{a,c\}$ is consistent; for example, it is satisfied
by setting $a$, $c$ and $m$ to true and all other variables to false.

The explanation $E'' \explain m$ is also in $\summarize(G,R)$. The clause $ab
\rightarrow m$ allows entailing $m$ since $E''$ contains both $a$ and $b$. The
consistency of $G \cup E''$ is proved by showing a model that satisfies it.
This model sets $x_i$ to true if $E''$ contains $x_i^p$ and to false if $E''$
contains $x_i^n$. It also sets $a$, $b$ and $m$ to true and all other variables
to false.

Since forgetting contains both $E \explain m$ and $E'' \explain m$ for every
evaluation of $X$, overreaching monotony requires $E' \explain m$ as well for
every evaluation of $X$. This is proved to be the case if the QBF is false.
This way, overreaching monotony is violated when the QBF is true.

Two clauses of $G \cup E'$ may entail $m$:
{} $ac \rightarrow m$
and
{} $FM = F \vee \neg a \vee c \vee
{}       (\neg y_1^p \neg y_1^n) \vee ... \vee (\neg y_n^p \neg y_n^n) \vee m$.
The clause $ab \rightarrow m$ does not entail $m$ since one of its
preconditions is $b$, which occurs positive neither in $G$ nor in $E'$.

The QBF is false if $\exists Y \forall Z . \neg F$ holds for all evaluations of
$X$. Each such evaluation corresponds to a set of hypotheses $E'$ containing
$x_i^p$ if the evaluation sets $x_i$ to true and $x_i^n$ otherwise. This way,
either $x_i^p$ or $x_i^n$ is in $E'$ for every index $i$. Since $E'$ also
contains $a$, the clause
{} $XC = a (x_1^p \vee x_1^n) \ldots (x_n^p \vee x_n^n) \rightarrow \neg c$
makes $G \cup E'$ entail $\neg c$. Therefore, adding $c$ to $E'$ violates
consistency.

The clauses
{} $XPX = \{x_i^p \rightarrow x_i \mid 1 \leq i \leq n\}$ and
{} $XNX = \{x_i^n \rightarrow \neg x_i \mid 1 \leq i \leq n\}$
force the values of $x_i$ as in the evaluation. The same goes for $Y$: for each
of its evaluations, the corresponding hypotheses $y_i^p$ and $y_i^n$ make $G$
entail the value of $y_i$. Adding these hypotheses to $E'$ makes it entail
$\neg F$ since $\forall Z . \neg F$ holds for these evaluations of $X$ and $Y$.
As a result,
{} $FM =
{}	F \vee \neg a \vee c \vee
{}	(\neg y_1^p \wedge \neg y_1^n) \vee \cdots \vee
{}	(\neg y_n^p \wedge \neg y_n^n) \vee
{}	m$
and $E'$ entail $m$. This proves that if the QBF is false, then every such $E'$
explains $m$, satisfying overreaching monotony. The converse is also the case.
If the QBF is true, the precondition of this clause is falsified by a value of
$Z$. The variable $m$ is not entailed. Overreaching monotony is violated.

This proves that overreaching monotony is violated if the QBF is true. The
proof is completed by showing that if the QBF is false, $\summarize(G,R)$ is
supported by a formula. Lemma~\ref{satisfies-monotony} then proves that
$\summarize(G,R)$ satisfies overreaching monotony.

The set of explanations $\summarize(G,R)$ always comprises
{} $a \explain m$,
{} $ab \explain m$,
and
{} $abE \explain m$
for all sets $E$ including some $x_i^p$ and $x_i^n$ but not both for the same
index. It also comprises $aE \explain m$ for all $E$ including some $x_i^p$ and
$x_i^n$ but not both for the same index, and not for all indexes $i$. Since the
QBF is false, it also includes the explanations $aE \explain m$ where $E$
contains either $x_i^p$ or $x_i^n$ but not both for all indexes $i$. Such
explanations are supported by the following formula.

\[
\{a \rightarrow m\} \cup
\{x_i^n x_i^p \rightarrow \bot \mid 1 \leq i \leq n\}
\]

This provides an alternative proof of the hardness of the existence of a
formula supporting $\summarize(G,R)$.~\qed

\subsection{Complexity characterization}

The following theorem sums up what was proved about the complexity of the
problem of the existence of a formula supporting forgetting.

\begin{theorem}
\label{complexity}

The following problems are \P{3}-complete:

\begin{itemize}

\item checking whether $\summarize(F,R)$ satisfies the conjunctive property;
\item checking whether $\summarize(F,R)$ satisfies overreaching monotony;
\item checking whether $\summarize(F,R)$ is supported by some formula.

\end{itemize}

\end{theorem}

\proof Consequence of
Lemma~\ref{membership-and},
Lemma~\ref{membership-monot},
Lemma~\ref{hard-and} and
Lemma~\ref{hard-monot}.~\qed

\section{Default logic}
\label{section-default}

\draft

\begin{itemize}

\item abduction by a default theory <D,W> supports E=>M if and only if M is
entailed by some extension of <D,WuE>; this is the credulous semantics because
the skeptical one satisfies the conjunctive condition, which may be falsified
by forgetting; this mechanism supports forgetting from every propositional
formula

\item the way it does is purely theoretical because it builds a default from
a set of explanations, which may be exponentially many in the size of the
formula; some possible ideas to improve it are presented

\end{itemize}

\enddraft

Forgetting may not be supported by any propositional formula. The keys of this
sentence are ``supported'' and ``propositional''. Forgetting always exists: it
is a set of explanations. It could be just stored as such, if not for its sheer
size: a tiny formula may support a myriad of explanations. This is a primary
reason for finding a propositional formula supporting it, because that formula
may have a reasonable size. Another is that a formula may provide insight of
what these explanations collectively indicate.

For some logics, forgetting can always be expressed in the logic
itself~\cite{lang-etal-03}. For some others, it may
not~\cite{gonc-etal-16,fang-etal-19,zhao-etal-20}. In such cases, a solution is
to switch to a more powerful logic~\cite{gonc-etal-16,zhao-etal-20}. For
example, strongly persistent forgetting is not always possible in logic
programming~\cite{gonc-etal-16-a,gonc-etal-20}, but it is extending the
language with forks~\cite{agua-etal-19}.


For propositional logic, the first choice are other logics made from simple
propositions, no objects or functions. This rules out first-order and
description logics, for example. Obvious candidates are modal logic and
nonmonotonic logics.

Default logic is an example. Its base language is that of propositional logic.
It defines entailment, which allows to derive consistency, the building blocks
of abduction.

A Reiter default theory~\cite{reit-80} is a pair $\l D,W \r$ where $W$ is a
propositional formula and $D$ a set of rules of the form $\frac{A:B}{C}$, which
means that $C$ is a typical consequence of $B$ when $A$ holds. If so, the
default is called applicable, and the addition of its consequent is the result
of applying it. The justification of this default is $B$, its consequent is
$C$. The semantics of default logic define its extensions, each being a set of
propositional formulae.

More details are in surveys of the topic~\cite{besn-89,anto-99}.

An explanation $E \explain M$ is supported if the default theory $\l D, W \cup
E\r$ is consistent and entails $M$~\cite{eite-etal-97,tomp-03}. The question
is: given a set of explanations $S$, is there any default theory $\l D,W \r$
that supports them, and them only? More specifically, if $S$ is the result of
forgetting, is there a theory supporting its explanations and no other?

The definition of forgetting from a default theory requires some additional
specifications. The same default theory may have multiple extensions. Some
extensions may entail $M$ and some may not. In such cases, does the default
theory entail $M$? Accepting $M$ as a consequence only if entailed by all
extensions satisfies the conjunctive condition: if $a$ and $b$ are consequences
of all extensions, so is $a \wedge b$. No default theory supports any set of
explanations violating the conjunctive condition, which forgetting may produce.
Accepting $M$ as a consequence if entailed by some extensions does not suffer
from this limitation: $a$ may be entailed only by extensions that do not entail
$b$ and vice versa; none entail $a \wedge b$.

Support for an explanation means that the default theory entails the
manifestation when added the hypotheses.

\begin{definition}

The explanation $E \explain M$ is supported by $\l D,W \r$ if $\l D,W \cup E\r$
has at least a consistent extension where $M$ holds.

\end{definition}

A simple case demonstrates that default logic supports sets of explanations
that propositional logic does not. The default theory $\l D,\emptyset \r$ that
follows supports $\{a \explain x, a \explain y\}$, which violates the
conjunctive condition because it does not contain $a \explain x \wedge y$.

\ttytex{
\begin{eqnarray*}
D &=& \left\{
	\frac{a:x \wedge \neg y}{x \wedge \neg y},
	\frac{a:\neg x \wedge y}{\neg x \wedge y}
\right\}
\\
W &=& \emptyset
\end{eqnarray*}
}{
      a:x-y  a:-xy
D = { -----, ----- }
       x-y    -xy
W = 0
}

The two extensions of $\l D,\emptyset \cup \{a\}\r$ respectively entail $x
\wedge \neg y$ and $\neg x \wedge y$. One entails $x$, one entails $y$, none
entail $x \wedge y$. The only supported explanations are $a \explain x$ and $a
\explain y$. No propositional theory supports them without also supporting $a
\explain x \wedge y$. The conclusion is that default logic support some sets of
explanations that propositional logics does not.

The question is whether forgetting from a propositional formula is always
supported by a default theory.

The answer is: yes, in theory.

The principle is demonstrated by the example: every explanation $E \explain M$
turns into a default, which requires $E$ and produces $M$ but none of its
supersets.

The resulting theory supports exactly the explanations that result from
forgetting. This claim is proved in three steps:

\begin{itemize}

\item a condition on sets of explanations is defined;

\item it is proved to be equivalent to the set being the result of forgetting
from a propositional formula;

\item it is proved to be equivalent to the set being supported by a default
theory.

\end{itemize}

The conclusion is that a set of explanations is the result of forgetting from a
propositional theory if and only if it is supported by a default theory. In
short, forgetting from a propositional theory results in a default theory.

\begin{definition}[Consequential monotony]
\label{consequential}

A set of explanations $S$ satisfies {\em consequential monotony} if it contains
$E \explain M'$ whenever it contains $E \explain M$ with $M' \subseteq M$.

\end{definition}

Consequential monotony characterizes forgetting. A set of explanations is the
result of forgetting some variables from some formulae if and only if it
satisfies consequential monotony. The first step of the proof is that
forgetting satisfies consequential monotony.

\begin{lemma}
\label{consequential-always}

For every formula $F$ and set of variables $R$, consequential monotony is
satisfied by $\focus(F,R)$.

\end{lemma}

\proof The premises are $E \explain M' \in \focus(F,R)$ and $M \subseteq M'$.
The conclusion is $E \explain M \in \focus(F,R)$.
The first premise $E \explain M' \in \focus(F,R)$ is defined as the existence
of $E'$ and $M''$ such that $F \models EE' \explain M'M''$. This is in turn
defined as $F \cup E \cup E'$ being consistent and entailing $M' \cup M''$. The
latter $F \cup E \cup E' \models M' \cup M''$ implies $F \cup E \cup E' \models
M \cup M''$ since $M \subseteq M'$. With the consistency of $F \cup E \cup E'$,
this entailment defines $F \models EE' \explain MM''$. This is the case for
some $E'$ and $M''$.
This is the definition of $E \explain M \in \focus(F,R)$.~\qed

The second step of the proof is that if a set of explanations satisfies
consequential monotony, it is a result of forgetting.

\begin{lemma}
\label{consequential-enough}

If a set of explanations $S$ over disjoint hypotheses $I$ and manifestations
$C$ satisfies consequential monotony then there exists a formula $F$ such that
$S = \focus(F,I \cup C)$.

\end{lemma}

\proof A new hypothesis is created for each explanation $E \explain M$ in $S$.
To this aim, $S$ is assumed enumerated: $S = \{E_i \explain M_i \mid 1 \leq i
\leq n\}$. The new hypotheses are $a_1,\ldots,a_n$. The formula $F$ comprises
the following clauses for each pair of indices $i$ and $j \not= i$:

\ttytex{
\begin{eqnarray*}
&& E_ia_i \rightarrow M_i			\\
&& a_i a_j \rightarrow \bot			\\
&& \{a_i e \rightarrow \bot \mid e \in I \backslash E_i\}
\end{eqnarray*}
}{
Eiai->Mi
aiaj->-
{aie->- | e c I-Ei}
}

These clauses make $F$ support $E_i \cup \{a_i\} \explain M_i$ for every
explanation $E_i \explain M_i$ of $S$. Forgetting $a_i$ turns it into $E_i
\explain M_i$, as required. This proves that every explanation of $S$ is
in $\focus(F,I \cup C)$.

The rest of the proof shows that the explanations that are not in $S$ are not
supported.

This is proved by contradiction: some explanation $E \explain M$ not in $S$ is
assumed to be in $\focus(F,I \cup C)$. By definition, this is only possible if
$F \cup E \cup D$ is consistent and entails $M$ for some $D \subseteq
\{a_1,\ldots,a_n\}$.

If $D$ is empty, $F \cup E \cup D$ is satisfied by a model that sets all
variables $a_i$ and $C$ to false, since all clauses of $F$ contain a negative
occurrence of a variable $a_i$. This model falsifies $M$, contradicting the
assumption that $F \cup E \cup D$ entails $M$.

If $D$ contains two variables $a_i$ and $a_j$, it falsifies the clause $a_i a_j
\rightarrow \bot$, contradicting the assumed consistency of $F \cup E \cup D$.

The conclusion is that $D$ contains exactly one variable $a_i$. Therefore, $F
\cup E \cup D$ is $F \cup E \cup \{a_i\}$.

The unions $F \cup E \cup \{a_i\}$ and $F \cup E \cup \{a_i\} \cup \neg M$
contain only negative occurrences of the variables $a_j$ with $j \not= i$.
Removing the clauses containing them does not affect satisfiability. The
remaining clauses of $F$ are
{} $E_i a_i \rightarrow M_i$ and
{} $a_i e \rightarrow \bot$ for every $e \in I \backslash E_i$.
That $F \cup E \cup \{a_i\}$ is consistent and entails $M$ simplify as follows.

\begin{eqnarray*}
\{E_i a_i \rightarrow M_i\}
\cup 
\{a_i e \rightarrow \bot \mid e \in I \backslash E_i\} \cup
E \cup \{a_i\}
& \not\models & \bot
\\
\{E_i a_i \rightarrow M_i\}
\cup 
\{a_i e \rightarrow \bot \mid e \in I \backslash E_i\} \cup
E \cup \{a_i\}
& \models & M
\end{eqnarray*}

If a variable $e$ of $E$ is not in $E_i$ then it is in $I \backslash E_i$. As a
result, the premise of the first entailment contains $a_i$, $e$ and $a_i e
\rightarrow \bot$, contradicting its consistency. The conclusion is that all
variables of $E$ are in $E_i$, which is the same as $E \subseteq E_i$.

If a variable $e$ of $E_i$ is not in $E$, the second entailment is contradicted
by the model that sets all variables of $\{a_i\} \cup E$ to true and all others
to false, including $e$. This model satisfies $E_i a_i \rightarrow M_i$ because
it sets $e$ to false. It satisfies every clause $a_i e \rightarrow \bot$
because every $e$ in $I \backslash E_i$ is in $I \backslash E$ since $E
\subseteq E_i$, and is therefore set to false by the model. The same model
falsifies $M$, contradicting the second entailment above.

The conclusion is that $E = E_i$. It turns the second entailment above into the
following one.

\[
\{E_i a_i \rightarrow M_i\}
\cup 
\{a_i e \rightarrow \bot \mid e \in I \backslash E_i\} \cup
E_i \cup \{a_i\}
\models M
\]

Since the premise contains $a_i$ and every $e \in E_i$, all negative
occurrences of these variables can be removed from the clauses where they
occur.

\[
M_i
\cup 
\{\neg e \mid e \in I \backslash E_i\} \cup
E_i \cup \{a_i\}
\models M
\]

Because of the separation of the variables, this is the same as $M_i \models
M$, which is the same as $M \subseteq M_i$. Since $E_i \explain M_i$ is in $S$,
by consequential monotony also $E_i \explain M$ is in $S$. This implies that
$E \explain M$ is in $S$ since $E = E_i$. Contradiction with the assumption
that $E \explain M$ is not in $S$ is reached.~\qed

The two lemmas prove that consequential monotony characterizes forgetting.

\begin{theorem}
\label{consequential-is-forget}

A set of explanations $S$ over disjoint hypotheses $I$ and manifestations $C$
satisfies consequential monotony if and only if there exist a formula $F$ such
that $S = \focus(F,I \cup C)$.

\end{theorem}

\proof Lemma~\ref{consequential-always} states that $\focus(F,R)$ satisfies
consequential monotony for every set of variables $R$, including $I \cup C$ for
whichever disjoint sets $I$ and $C$. Lemma~\ref{consequential-enough} proves
the other direction.~\qed

Consequential monotony also equates the existence of a default theory
supporting the set of explanations.

\begin{theorem}
\label{forget-as-default}

A set of explanations $S$ over disjoint hypotheses and manifestations satisfies
consequential monotony if and only if it supported by a default theory.

%

\end{theorem}

\proof The claim comprises two parts: first, if a default theory supports $S$,
then $S$ satisfies consequential monotony; second, if $S$ satisfies
consequential monotony, a default theory supports it.

The first part holds because the explanations supported by an arbitrary default
theory satisfy consequential monotony. The premise is that an explanation $E
\explain M$ is supported by a default theory $\l D,W \r$. This is defined as
the existence of a consistent extension of $\l D,W \cup E\r$ that entails $M$.
An extension is just a propositional formula. Since it entails $M$, it also
entails every subset $M' \subseteq M$. As a result, the default theory supports
$E \explain M'$, as required.

The second part of the claim is that every set of explanations $S$ over
disjoint hypotheses and manifestations satisfying consequential monotony is
supported by some default theory. This default theory is $\l D,\emptyset \r$,
comprising the following defaults.

\ttytex{
\[
D = \left\{\left.
\frac
{E:E \wedge \neg (I \backslash E) \wedge M \wedge \neg (C \backslash M)}
  {E \wedge \neg (I \backslash E) \wedge M \wedge \neg (C \backslash M)}
\right| E \explain M \in S
\right\}
\]
}{
      E:E -(I-E) M -(C-M)
D = { ------------------- | E=>M in S }
       E -(I-E) M -(C-M)
}

The claim is that $E \explain M$ is in $S$ if and only if some consistent
extension of $\l D, \emptyset \cup E\r$ entails $M$.

The inclusion of $E$ in the justification and consequent of the default is
redundant, but facilitates some parts of the proof.

Since the defaults of $D$ are normal and the background theory $\emptyset \cup
E$ is consistent since $E$ is a set of positive literals, $\l D,\emptyset \cup
E\r$ always has at least an extension, and all its extensions are consistent.

A preliminary result is that no two defaults can be applied together. Since $S$
is a set and not a multiset, its explanations $E \explain M$ differ from each
other. Every two of them differ either on $E$ or on $M$: if $E \explain M$ and
$E' \explain M'$ are both in $S$, then either $E \not= E'$ or $M \not= M'$.
Four cases are possible:

\begin{itemize}

\item $E$ contains a hypothesis not in $E'$;

\item $E'$ contains a hypothesis not in $E$;

\item $M$ contains a manifestation not in $M'$;

\item $M'$ contains a manifestation not in $M$.

\end{itemize}

Only the first case is considered, the other three are similar due to the
symmetry of the defaults. Let $e \in E \backslash E'$. This variable belongs to
$E$, and therefore occurs positive in the justification and consequent of the
default of $E \explain M$; since it belongs to $E \backslash E'$, it belongs to
its superset $I \backslash E'$; as a result, it occurs negative in the
justification and consequent of the default of $E' \explain M'$. Applying the
default of $E \explain M$ results in the generation of $e$, which blocks the
application of the default of $E' \explain M'$. Applying the latter results in
the generation of $\neg e$, which blocks the application of the former.

This proves that no two defaults can be applied together.

The main claim can now be proved: $E \explain M$ is in $S$ if and only if a
consistent extension of $\l D,\emptyset \cup E\r$ entails $M$. Two cases are
considered: either $E \explain M$ is in $S$, or it is not. The claim is that an
extension of $\l D,\emptyset \cup E\r$ entails $M$ in the first case and no
extension entails $M$ in the second.

\begin{description}

\item[$E \explain M \in S$]

By construction, $D$ contains the default of $E \explain M$. Its premise is
$E$, which holds in $\l D,\emptyset \cup E\r$. Its consequent includes $E$
itself and $M$; this part is consistent with $\emptyset \cup E$ because all
these sets comprise positive literals only; it also includes $\neg (I
\backslash E)$ and $\neg (C \backslash M)$; these two sets comprise negative
literals, but they are disjoint from the positive literals $E \cup M$ because
of the separation between hypotheses and manifestations $I \cap C = \emptyset$.
A consequence of this consistency is that the default is applicable. Its
application blocks all other defaults. An extension is generated, and this
extension includes $M$.

\item[$E \explain M \not\in S$]

The claim that no extension of $\l D,\emptyset \cup E\r$ entails $M$ is proved
by contradiction: an extension entailing $M$ is assumed to exist.

As proved above, every extension is generated by the application of zero or one
default.

Applying zero defaults to $\l D,\emptyset \cup E\r$ adds nothing to the
background theory $\emptyset \cup E$. This set does not entail $M$ because of
the assumptions $M \not= \emptyset$ and $I \cap C = \emptyset$. This
contradicts the assumption that the extension entails $M$.

The other possibility is that the extension results from applying exactly one
default. By construction, every default comes from an explanation $E' \explain
M'$ of $S$. Since this explanation belongs to $S$, it is not the same as $E
\explain M$, which does not. Either $E' \not= E$ or $M' \not= M$.

Since the default of $E' \explain M'$ is applied to the background theory
$\emptyset \cup E$, its precondition is entailed and its justification is
consistent. The entailment $\emptyset \cup E \models E'$ implies $E' \subseteq
E$. The consistency of
{} $\emptyset \cup E \cup
{}  \{E' \wedge \neg (I \backslash E') \wedge
{}    M' \wedge \neg (C \backslash M')\}$
implies that $E$ does not contain any hypothesis in $I \backslash E'$. This
condition
{} $E \cap (I \backslash E') = \emptyset$
translates into
{} $E \subseteq E'$.
Since the converse also holds, the containment is actually an equality: $E =
E'$.

The extension generated by the application of this default is the deductive
closure of its consequent and the background theory. It was assumed to entail
$M$. This entailment is the same as
{} $\emptyset \cup E \cup
{}  \{E' \wedge \neg (I \backslash E') \wedge
{}    M' \wedge \neg (C \backslash M')\} \models M$.
Because of the separation of hypotheses and manifestations, this is the same as
{} $M' \wedge \neg (C \backslash M') \models M$,
which is also the same as $M' \models M$, or $M \subseteq M'$.

What proved so far is $E = E'$ and $M \subseteq M'$ for some explanation $E'
\explain M'$ of $S$. A consequence is that $S$ contains $E \explain M'$ with $M
\subseteq M'$. By consequential monotony, $S$ also contains $E \explain M$,
contrary to the assumption.

\end{description}~\qed

This proves that abductive forgetting in propositional logics is supported by
default logic abduction.

In theory.

In practice, the default theory that supports a set of explanations is nothing
more than the explanations themselves, each turned into a default. It gives no
intuition other than that, which is a problem if the aim of forgetting is to
provide a summary of knowledge. Even if it is not, it is a computational
drawback. Forgetting may generate many explanations even from a small formula.
A default theory that always contains a default for every explanation is as
large as the set of explanations itself. This size increase may be unavoidable
in the worst case, but should be avoided if possible.

Theory proves that every consequential monotonic set of explanations is
supported by a certain default theory, but that default theory may be too large
for practical purposes. At the same time, theory does not forbid smaller
default theories to support the same set.

A smaller default theory may exploit the background theory $W$. Instead of
$\emptyset$, the consequential forgetting of $F$ could be used instead.
Consequential forgetting retains all and only the implications $E \rightarrow
M$ that do not contain hypotheses to forget. The default theory $\l \emptyset,W
\r$ supports all explanations $E \explain M$ of this kind. It is however
incomplete, as forgetting also contains explanations $E \explain M$ that are
not supported by the original formula. An example is $a \explain x$ where the
original formula supports $ab \explain x$ instead and $b$ is forgotten.
Consequential forgetting does not turn $ab \rightarrow x$ into $a \rightarrow
b$. Therefore, it has to be enhanced to support $a \explain x$. A way to do
this could be by a default rule.

\[
\frac{E:M \wedge \neg E_1 \wedge \cdots \wedge \neg E_n}{M}
\]

The justification and the consequent are no longer the same. This default is
not normal. It could not, because adding $\neg E_i$ to the background theory
might have unwanted consequences. For example, it could entail the negation of
a manifestation in $M$. For this reason, $\neg E_i$ is only in the
justification, to block the application of this default without producing
consequences.

\section{New variables}
\label{section-variables}

\draft

\begin{itemize}

\item forgetting is not always supported by a formula because the forgotten
hypotheses may interact among them and with the remembered hypotheses

forgetting can be supported by somehow expressing these interactions

one way to do that is to introduce new variables, that are then forgotten

\item it is not necessarily the same as reintroducing the old variables; as an
example, one single new variable may allow and be necessary to support
forgetting an arbitrary large number of hypotheses

\item again for a single new variable: it satisfies the conjunctive condition;
therefore, it only allows supporting a set of explanations violating
overreaching monotony, but not the conjunctive condition

\item how to adapt G(S) to a single new variable; still has the property that
G(S) supports S if and only such a formula exists

\end{itemize}

\enddraft

Abductive forgetting from a propositional formula may produce a set of
explanations that is not supported by a propositional formula. It is supported
by a default theory, which is an extension of a propositional formula. Is there
any other solution, one that does not require an extended logic?

An analysis of what makes propositional logic fail at abductive forgetting
suggests it.

Abductive forgetting mainly turns explanations like $ab \explain m$ into $a
\explain m$. It also removes the explanations of the manifestations that
comprise or include manifestations to forget, depending on the definition, but
this is less of a problem. What makes forgetting difficult is the removal of
hypotheses from explanations.

The meaning of $ab \explain m$ is that that $m$ is explained by $ab$.
Similarly, $a \explain m$ means that $m$ is explained by $a$. At the level of
English sentences, ``$a$ and $b$ explain $m$'' and ``forget about $b$'' result
in ``$a$ explains $m$''. If the only allowed verb is ``explain'', this is the
best that can be said: ``$a$ explain $m$''. As a matter of fact, ``$a$ {\em
might} explain $m$'' would be better. In certain conditions, $a$ explains $m$.
The certain conditions are $b$. These conditions are neglected. They should,
because this is what forgetting is supposed to do: neglect the conditions to
forget.

Yet, these conditions may interact with each other and may interact with the
other hypotheses.

\begin{itemize}

\item an example of interaction between neglected conditions is
{} $F = \{ab \rightarrow m, ac \rightarrow m', abc \rightarrow \bot\}$;
forgetting $b$ and $c$ produces $a \explain m$ and $a \explain m'$: in certain
conditions ($b$), an explanation of $m$ is $a$; in certain other conditions
($c$), an explanation of $m'$ is $a$. Yet, these two certain conditions never
materialize together because of $abc \rightarrow \bot$;

\item an example of interaction between neglected conditions and hypotheses is
{} $F = \{ab \rightarrow m, abc \rightarrow \bot\}$;
forgetting $b$ only turns $ab \explain m$ into $a \explain m$; in certain
conditions ($b$), an explanation of $m$ is $a$; yet, these conditions
$b${\plural} prevent the hypothesis $c$ to be the case, excluding $ac$ as a
further explanation of $m$.

\end{itemize}

Forgetting may be supported by specifying the ``certain conditions'' that do
not materialize together, and the ones that prevent other hypotheses to
materialize.

A way to formalize this is by attacks like in argumentation
theory~\cite{dung-95}: $a \explain m$ and $a \explain m'$ attack each other,
meaning that the conditions that make $a$ an explanation of $m$ conflict with
the ones that makes it an explanation of $m'$; in the same way, $a \explain m$
attacks $c$, meaning that conditions that make $a$ an explanation of $m$
conflict with $c$.

A simpler solution is to use introduce new hypotheses.

This is always possible because the new hypotheses can just be the forgotten
ones. Yet, they might not. The ``certain conditions'' may be complicated but
what matters might be only that two of them conflict with each other. For
example, the original theory may entail $aC \rightarrow m$ and $aC' \rightarrow
m'$, where $C$ and $C'$ are complicated formulae that are not consistent with
each other. Forgetting their variables result in $a \explain m$ and $a \explain
m'$ only, without $a \explain mm'$. The same is the result of forgetting $b$
and $c$ from $\{ab \rightarrow m, ac \rightarrow m', bc \rightarrow \bot\}$.
Complicated conditions $C$ and $C'$ are turned into two simple hypotheses $b$
and $c$. A simple clause $bc \rightarrow \bot$ forbids them to be both true at
the same time.

Forgetting is supported by forgetting, of course. But may not only be supported
by forgetting the same variables from the same formula. It may be supported by
forgetting variables from a simpler formula. This is economy of concepts:
hypotheses are introduced only when they are necessary to support the
explanations resulting from forgetting. If a forgotten hypothesis does not
conflict with any other, it can be just removed without the need of introducing
any new one. In other cases, multiple forgotten hypotheses involved in complex
subformulae can be summarized with two new hypotheses, like in the example $b$
and $c$ take over $C$ and $C'$.

An extreme example shows that sometimes a single new hypothesis can replace
arbitrarily many forgotten ones.

\begin{eqnarray*}
F &=& \{
	a_ix_i \rightarrow m_i,
	a_ib_ix_i \rightarrow \bot,
	a_ib_i \rightarrow m'_i
\mid 1 \leq i \leq m
\}
\cup						\\
&&
\{
a_i a_j \rightarrow \bot ,~ a_i b_j \rightarrow \bot, a_i x_j \rightarrow \bot
\mid 
1 \leq i \leq m,
1 \leq j \leq m,
i \not= j
\}
\end{eqnarray*}

The explanations supported by this formula are $a_ix_i \explain m_i$ and
$a_ib_i \explain m_i'$ for every index $i$ between $1$ and $m$. Forgetting
$x_i$ turn them into $a_i \explain m_i$ and $a_ib_i \explain m_i'$, which are
not supported by any formula because they violate overreaching monotony. The
variables $x_i${\plural} need not be all different to produce them. A single
variable $x$ suffices.

\begin{eqnarray*}
F' &=& \{
	a_ix \rightarrow m_i,
	a_ib_ix \rightarrow \bot,
	a_ib_i \rightarrow m'_i
\mid 1 \leq i \leq m
\}
\cup						\\
&&
\{
a_i a_j \rightarrow \bot ,~ a_i b_j \rightarrow \bot
\mid
1 \leq i \leq m,
1 \leq j \leq m,
i \not= j
\}
\end{eqnarray*}

This is the case for an arbitrary large $m$: forgetting a single variable may
support forgetting arbitrarily many hypotheses, even when that violates
overreaching monotony and is therefore not supported by plain abduction without
forgetting.

Forgetting a single variable may support forgetting multiple ones. It may, but
it also may not. It depends on the formula and on the variables.

This is not merely a matter of numbers, of how many variables are forgotten;
what matters is the kind of supported explanations. Forgetting a single variable
may violate overreaching monotony, but never violates the conjunctive
condition. This is at the same time a blessing and a curse.

\begin{itemize}

\item When forgetting is originally done on a single variable, the result is at
least guaranteed to satisfy one of the conditions for being supported by a
propositional formula.

\item When trying to express the result of forgetting, a single additional
variable only helps when the conjunctive condition is satisfied.

\end{itemize}

Forgetting may in general violate the conjunctive condition. It does only when
forgetting at least two hypotheses.

\begin{theorem}

Forgetting a single hypothesis from a formula satisfies the conjunctive
condition.

\end{theorem}

\proof The claim is shown in reverse: if forgetting violates the conjunctive
condition, the forgotten hypotheses are at least two.

Forgetting violates the conjunctive condition when it contains $E \explain M_1$
and $E \explain M_2$ and not $E \explain M_1M_2$. This is the case when the
formula $F$ supports $EA \explain M_1$ and $EB \explain M_2$ for some sets of
hypotheses to forget $A$ and $B$, and it does not support $EC \explain M_1M_2$
for any set of hypotheses to forget $C$. This includes $C = A \cup B$: it does
not support $EAB \explain M_1M_2$.

Since $F$ supports $EA \explain M_1$ and $EB \explain M_2$, it entails $EA
\rightarrow M_1$ and $EB \rightarrow M_2$. As a result, it entails $EAB
\rightarrow M_1M_2$. If $F \cup E \cup A \cup B$ were consistent, then $F$
would support $EAB \explain M_1M_2$; it does not; therefore, $F \cup E \cup A
\cup B$ is inconsistent.

Since $F \cup E \cup A \cup B$ is inconsistent while its subset $F \cup E \cup
A$ is not, their difference $B$ contains at least a hypothesis that is not in
their intersection $F \cup E \cup A$, and is therefore not in $A$. For the same
reason, $A$ contains a hypothesis that is not in $B$. Since $A$ and $B$ are
sets of hypotheses to forget, these are at least two.~\qed

This result allows for a different tentative-supporting formula of a set of
explanations. The implication $E \rightarrow m$ is generated from $E \explain
m$ only if this explanation is not involved in a violation of overreaching
monotony. The other case is the absence of $E' \explain m$ and the presence of
$E'' \explain m'$ for some other sets of hypotheses $E'$ and $E''$ such that $E
\subseteq E' \subseteq E''$ and some other manifestation $m'$. The algorithm
generates $Ex \rightarrow m$ and $Ex \rightarrow \neg (E' \backslash E)$ then.
These two clauses support $E \explain m$ and block $E' \explain m$ while
allowing $E'' \explain m$ when forgetting $x$.

If forgetting violates overreaching monotony only in one case, the resulting
formula supports it. Otherwise, it may support it or not. If all violations are
somehow independent on each other, for example they are all about different
sets $E$, it works. Otherwise, it may not. It is only an attempt at supporting
the given set of explanations anyway, hence the name tentative-supporting
formula.

\section{Related work}
\label{section-related}

Many authors used forgetting as a way to find the explanation of a specific
manifestation. The underlying principle is that an explanation can be found by
negating the manifestations, conjoining the theory, forgetting everything but
the hypotheses and negating the result~\cite{lin-01}. This mechanism is applied
to propositional logic~\cite{lin-01}, description
logics~\cite{koop-schm-15-a,delp-schm-19,delp-schm-19-b,koop-20}, logic
programming~\cite{wern-13} and modal logics~\cite{feng-etal-18}.


Lobo and Uzc{\'{a}}tegui~\citeyear{lobo-uzca-97} characterized the logical
inference relation deriving from abduction. A formula is a consequence of
another if it is a consequence of all its explanations. Like $\explain$ in the
present article, this relation is based on the abductive explanations of a
fixed formula. Contrary to that, it satisfies the conjunctive condition by
construction, being based on logical inference. Expressing such a relation with
a cumulative model is akin to supporting $\explain$ by a formula: a binary
relation derived from abduction is expressed as something else. Yet, the binary
relations differ, as do their alternative expressions.



Pino-Perez and Uzc{\'{a}}tegui~\citeyear{pino-uzca-03} also define a binary
relation based on abduction, but theirs is almost identical to $\explain$ apart
from the order of their arguments. The only semantical difference is that its
second argument is always an explanation of the first. This prevents encoding
the explanations of forgetting, which may not be the explanations of a formula.
While in the present article $\explain$ can or cannot not be encoded as a
formula to abduce from, their relation can or cannot be defined in terms of an
ordering among formulae. While the two relations are mostly the same, the way
they are expressed differ.

%


Beierle etal.~\cite{beie-etal-19} define a very general notion of forgetting
from a doxastic state, which they instantiate to various operations of belief
change. This specific instantiation differs from abduction, which however fits
into the general framework: an abduction frame supporting an explanation is a
specific case of a doxastic state (the abduction frame) having a consequence
(the explanation). This view allows a comparison between abductive forgetting
and the various forms of abstract forgetting, the closest ones being
abstraction, marginalization and focusing. Abstraction requires both $a
\explain xy$ and $a \explain x \neg y$ to infer $a \explain x$, differently
from forgetting in abduction. Instead, abductive forgetting matches
marginalization, the reduction of language. The specialization of
marginalization where the reduced alphabet is derived from the doxastic state
is called focusing.

What to forget is indeed typically a given in logical forgetting. The data is:
some kind of formula and some variables to forget. These variables are fixed.
This needs not to be the case. For example, when forgetting is used for
privacy, they are the facts not to disclose. However, they may not be enough.
For example, privacy preservation requires not to indirectly disclose
information. Address and age may allow finding names. GPS locations and times
imply addresses. Incomes suggest ages. Removing names and leaving related data
may be useless. The related data is to be canceled as well. This connects
forgetting with the logics of relevance~\cite{ande-beln-75,bimb-dunn-17}.


\section{Conclusions}
\label{section-conclusions}


Abductive forgetting having two different definitions come from its different
applications: concentrating on a topic and summarizing it. Multiple
definitions, based on different requisites, are common in forgetting in
formalisms other than propositional logics~\cite{gonc-etal-21,eite-kern-19}.

Whether abductive forgetting is supported by a propositional formula is not
only a matter of representation. It tells something about the scenario the
formula represents. Forgetting A turns ``AB explain C'' into ``B explain C'',
meaning: ``B may explain C'' or better ``in certain conditions not of interest,
B explain C''. Forgetting being supported by a formula means that the
conditions for different explanations do not interact in a relevant way. For
example, they do not conflict. They can really be neglected, because they do
not influence what forgetting maintains. For a given formula, this may happen
when forgetting certain variables but not others. The former are ``fully
forgettable'' because neglecting them does not introduce any complication. The
latter are not fully forgettable, as the result of forgetting is still affected
by them. A similar observation emerged in Answer Set Programming;
Aguado~et~al.~\citeyear{agua-etal-19} wrote: ``In practice, this means that
auxiliary atoms in ASP are more than ``just'' auxiliary, as they allow one to
represent problems that cannot be captured without them.''

The absence of a formula supporting forgetting may be a red flag in certain
applications. A professor summarizing a topic may conclude that the result is
too complicated. Too complicated to be even expressible in the same logic of
the whole topic. Some of the removed elements are too important. Some of the
maintained elements are too marginal. The alternative of presenting every
single explanation one by one may not be much of a summarization. The variables
to forget are to be changed.


Many questions are open.

\nojournal

The complexity of the problem is still open for the case of minimal
explanations.

\endnojournal

The algorithm that produces a formula that supports forgetting if any is
exploited in a theoretical context, for proving the necessary and sufficient
condition. Its practical application is limited to small formulae, since it is
exhaustive: it reads the set of all explanations of forgetting, which is in
general exponentially larger than the original formula. For large formulae, its
large running time prevents its use. A better choice would be to start by
consequentially forgetting, adding clauses only when necessary. The first step
is correct because forgetting comprises necessary clauses only: every entailed
clauses $E \rightarrow M$ and $E \rightarrow \bot$ made of variables to
remember only is needed to ensure the survival or removal of the explanation $E
\explain M$. While this first step is correct, it is not complete. Forgetting
may produce other explanations. The clauses needed to ensure need to be added
in a second step.

A related question is whether a formula supporting forgetting not only exists,
but is also of reasonable size. Forgetting always exists since it is defined as
a set of explanations, which may however comprise many explanations. If the
formula supporting it is similarly sized, it does not offer any
benefit~\cite{libe-24-a,libe-24-b}.

The tentative-supporting formula $G(S)$ looks unique in the way it supports a
given set of explanations. It is not syntactically minimal, as it may for
example contain both $E \rightarrow m$ and $E' \rightarrow m$ with $E \subset
E'$. It is however minimal in the sense that it only contains clauses that are
strictly necessary to support the given set of explanations $S$. Semantically,
it contains as many models as possible, among the formulae that have $S$ as
their supported explanations.

Other formulae supporting the same explanations may contain other clauses $E
\rightarrow \bot$. If no subset of $E$ explain anything, such a clause is not
mandatory. If no superset of $E$ explain anything, it is permitted. When both
are the case, the clause can be entailed or not. This looks like the only way
formulae supporting the same explanations may significantly differ on the
hypotheses and manifestations.


Default logic always expresses forgetting, but only with Reiter-style defaults.
The proof employs a theory that only comprises normal defaults and is therefore
uncontroversial~\cite{libe-05-b}. As a result, the proof extends to all
consonant default logics~\cite{libe-05-b} such as justified~\cite{luka-88},
constrained~\cite{scha-92,delg-scha-jack-94} and rational~\cite{miki-trus-95}.
Yet, these normal defaults are not prerequisite-free. It is therefore an open
question whether the proof extends to other forms of defaults such as
Poole's~\cite{pool-88}, Brewka's~\cite{brew-89} and Nebels's~\cite{nebe-91}.
Yet, nothing prevent them to express forgetting in another way.

Default logic always expresses forgetting, but other extensions of
propositional logic may do that. Yet, many logics have the conjunctive
property: they entail a conjunction if and only if they entail each of its
parts. Forgetting sometimes requires this not to be the case. The cases where
forgetting is not expressed by a formula suggests one: the problem is with the
interaction between the forgotten conditions of an explanation and the other
maintained explanations; that something invalidates something else is expressed
by arguments~\cite{dung-95,baum-etal-20,baum-bert-22}.

\nojournal
Introducing new variables to support forgetting makes sense only when they are
as few as possible. The principle is simplicity: the goal of forgetting is to
concentrate only on a set of variables; adding new ones may be a necessity, but
they are unwanted. Adding many complicates the result, when the aim of
forgetting is to simplify. Reducing the new variables as much as possible is
an open question.
\endnojournal

Explanations and manifestations are always sets of variables in this article.
They are positive literals. In general, they can be
formulae~\cite{lobo-uzca-97}. Extending the results to explanations and
manifestations that are formulae and employing complex preference
orderings~\cite{pino-uzca-03,delg-etal-04} is a further direction of study.

Many logics suffer from not being able to represent the result of
forgetting~\cite{lin-reit-94,wang-etal-09,gonc-etal-16-a,fang-etal-19}. A
workaround is not to represent the result of forgetting at all: an explanation
is searching from the original abduction frame and the variables to forget.
While the complexity of doing this has not been investigated yet, a preliminary
observation suggests it may not be that convenient. The definition of abductive
forgetting involves a quantifier alternation, making the search for an
explanation solution hard. If a formula represents forgetting, abduction
becomes relatively easy~\cite{eite-gott-95-a}, with the benefit that many
manifestations can be explained efficiently from it.

\draft
\section{other}

\v

spiegazioni
-----------

spiegazioni aggiuntive, possono servire per rispondere ai revisori

- forget e partial observability

  la prima definizione di forget e' collegata a partial observability, ma non
  e' solo partial observability; perche' oltre alla osservabilita' parziale
  delle manifestazioni anche le cause che interessano vengono limitate; e
  infatti nello studio successivo delle condizioni e della complessita' e'
  emerso che quella e' la parte difficile del problema

- le spiegazioni di una formula e del forget sono diverse:

  . le spiegazioni della formula sono condizioni sufficenti a spiegare le
    manifestazioni: quando E e' vero, lo e' anche M
  . le spiegazioni del forget sono precondizioni parziali: in generale, non
    basta che E sia vero perche' lo sia anche M; possono servire altre
    precondizioni, ma sono da dimenticare

  e' un problema se lo scopo e' rendere vero M, perche' la E del forget non
  basta; ma di solito lo scopo dell'abduzione non e' questo; M e' noto essere
  vero, e abduzione viene usata per sapere quali fra le cause possibili sono
  vere; dato che M e' vero, potrebbe esserlo anche E; se una parte di E non
  interessa, si puo' trascurare

  e' vero che in realta' E da sola non spiega M, ma in ogni caso E sarebbe
  comunque solo una spiegazione possibile; anche se non si trascura niente,
  puo' semplicemente non essere la spiegazione giusta, il motivo per cui M e'
  vero

  anche se il problema fosse quello di rendere vero M, in ogni caso
  dimenticarsi di alcune ipotesi puo' voler dire che non sono controllabili (e
  allora occorre fare forget universale invece che esistenziale) ma puo' anche
  voler dire che tanto sono cosi' facili da controllare che si possono
  trascurare

- la non esistenza del forget non e' sempre un problema

  in altre situazioni la inesistenza del forget viene considerata un problema,
  ma qui si vede che e' un valore; basta guardare l'esempio: quando si parla di
  a, x e y non ci si puo' dimenticare di b e c, perche' la spiegabilita' e'
  strettamente intercorrelata a queste; non e' piu' un caso di "a e altro che
  non interessa spiega x" e "a e altro che non interessa spiega y"; l'altro
  questa volta non puo' non essere di interesse, visto che la presenza e la
  assenza di b e c sono centrali per spiegare o non spiegare x e y; queste due
  variabili non possono essere dimenticate

  esempio piu' generale: riassunto del XX secolo, ma dimenticando di WW2,
  oppure del comunismo

  se non esiste una formula che esprime il forget, allora l'insieme delle
  variabili si puo' considerare una suddivisione non ragionevole rispetto alla
  formula, perche' non tiene insieme argomenti correlati; o almeno non e' una
  divisione ideale; questo porta alla definizione di "abductively isolable" o
  "abductively separable"

espansioni possibili
--------------------

- minimalita' (minimal.tex, tolta dalla versione su rivista)

- hardness nel caso di spiegazioni minime (minimal.tex)

- il discorso su dimenticare nuove variabili e sulle condizioni aggiuntive
  suggerisce un modo di esprimere le relazioni aggiuntive, che e' quello di
  argumentations, in cui un argomento ne puo' attaccare un altro; va pero'
  detto che li' la cosa e' in genere fatta in modo esplicito, quindi in questo
  caso esponenziale

- miglioramenti all'algoritmo

  quello attuale itera sulle spiegazioni E=>M da realizzare, che possono essere
  in numero esponenziale; dovrebbe almeno partire dalla formula F e dalle
  variabili da ricordare R, e non iterare sempre tutte le spiegazioni

  si potrebbe partire dal forget proposizionale della formula, che tanto deve
  essere implicato; si verifica in modo non esaustivo se esprime esattamente le
  spiegazioni; in caso contrario, aggiunge la clausole E->m e E->- che serve

- complessita' del problema di esprimere forget in spazio dato

- caso di ipotesi e manifestazioni intersecanti:

  . davvero e' un problema nel caso di tutte le spiegazioni o quelle minime
    per contenimento? sembrerebbe che si tratti solo di estendere il lemma
    body-contains al caso intersecante, ma non e' stato dimostrato ne' che si
    possa fare ne' che poi tutto il resto funzioni
  . ma piu' in generale, esiste una condizione su <= che dice quando si puo'
    usare G(S) definito come in questo articolo e quando no?
  . funziona la soluzione di aggiungere prima le clausole E->m e poi usare
    queste invece delle spiegazioni E=>M nella regola per aggiungere E->-?
  . questa soluzione richede di seguire un ordine specifico?
  . funziona il modo di aggiungere le clausole E->- ogni volta che e'
    possibile, invece che quando e' strettamente neceessario?

- G(S) e' unico, in qualche senso?

  per come e' costruita, sembra essere unico rispetto all'insieme S

  non e' sintatticamente minimo dato che puo' contenere clausole E->m e E'->m
  con EcE'; succede sempre, a meno che E' non debba essere inconsistente

  e' una formula sulle sole ipotesi e manifestazioni da ricordare R\&(IuC); fra
  queste, e' minima nel senso che e' quella che pone meno vincoli, dato che
  aggiunge clausole solo quando e' strettamente necessario; quindi e' quella
  con piu' modelli possibili

  semanticamente, la clausola E->- potrebbe non essere necessaria ma nemmeno
  vietata; in questo caso, G(S) non la contiene ma aggiungerla non cambia
  niente

  questo si ricollega al problema se e' possibile costruirne una formula simile
  a G(S) ma che contiene tutte le E->- possibili invece di solo quelle
  necessarie; era la formula proposta per risolvere al problema che si verifica
  quando ipotesi e manifestazioni intersecano

  la formula G(S) non e' unica, ma le formule su R\&IuC che supportano S
  dovrebbero essere tutte simili:
  . formule su IuC, dato che ogni queste sono le variabili indispensabili
  . implicano E->m e E->- ogni volta che G(S) lo fa
  . se G(S)uE e' consistente, allora le clausole E->m implicate sono le stesse

  sembra che l'unico caso in cui due teorie semanticamente diverse possono
  essere uguali in abduzione e' sulle spiegazioni E=>m non in S che possono
  venire escluse in entrambi i modi, sia per inconsistenza che per non
  implicazione; quindi E=>m non e' in S e al tempo stesso E->m non e'
  obbligatorio mentre l'inconsistenza di E e' consentita
  - E=>m non e' in S
  - una delle due:
    . E->m non e' obbligatorio
    . E->- e' consentito

  la prima cosa implica che S non contiene nessuna spiegazione E'->m con E'c=E;
  la seconda implica che S non contiene spiegazioni E'=>m' con Ec=E'; se tutte
  e due le cose sono vere, G(S) non contiene E->- ma volendo si puo' anche
  aggiungere

  ac=>x
  bc=>y
  abc non spiega niente
  ab non spiega niente

  quindi abc->- e' obbligatorio; ma ab->- non lo e', perche' comunque ab non
  implica niente; questo insieme di spiegazioni e' supportato da due formule
  semanticamente diverse:

  {ac->x, bc->y, abc->-}
  {ac->x, bc->y, ab->-}

  sono diverse perche' la prima e' consistente con ab-c; la seconda no

  altro esempio:

  ab=>x
  c=>y
  ac non spiega niente
  abc non spiega niente

  la clausola ac->- la posso mettere o meno, perche' tanto nessun sottoinsieme
  di ac spiega niente; e' vero che ab=>x, ma poi tanto abc va bloccato in ogni
  caso; quindi nessun sottoinsieme di E e nessun sovrainsieme deve spiegare
  niente; posso avere E'E''=>m', ma allora EE''=>m' non ci puo' essere

  in altre parole, le singole spiegazioni di E o anche i sottoinsiemi possono
  essere contenuti strettamente in altre spiegazioni; ma i sottoinsiemi e i
  sovrainsiemi di E non possono essere essi stessi spiegazioni:

    E'=>m non e' in S per nessun E' tale che E'c=E oppure Ec=E'

  se questo e' vero, allora G(S) e' abduttivamente equivalente a G(S)u{E->-};
  in caso contrario:

    - se esiste un sottoinsieme di E che spiega qualcosa, allora deve
      implicarlo; quindi anche anche E deve implicarlo, per cui E->- e'
      necessario e non opzionale

    - se esiste un sovrainsieme di E che spiega qualcosa, allora non puo'
      essere inconsistente; quindi nemmeno E, per cui E->- e' vietato e non
      opzionale

  questo dimostra che se la condizione e' falsa allora E->- e' necessario
  oppure vietato, quindi non opzionale; va trovato il modo di dimostrare che

  - se la condizione e' vera allora E->- si puo' aggiungere o meno senza altri
    effetti

  - questo e' l'unico caso di clausole opzionali; notare che fra le clausole
    che posso aggiungere ci sono anche clausole che contengono per esempio solo
    letterali positivi oppure piu' letterali positivi

- esistono altre logiche che non rispettano la regola (and)?

- volendo e' anche possibile combinare piu' cose, richiedendo per esempio che
  il forget mantenga le spiegazioni abduttive anche a fronte di modifiche;
  sarebbe una combinazione di forget abduttivo e revisivo

- potrebbe avere senso fare un forget misto consequenziale/consistenziale a
  seconda delle variabili da dimenticare; verificare se e' uguale al risultato
  di fare le due cose in un ordine qualsiasi

lavori correlati
----------------

quelli rilevanti sono gia' nella sezione di lavori correlati

qui maggiori dettagli

- il forget con simboli variabili e' comunque un forget universale sulle
  conseguenze; e' una combinazione di forget di letterali e variabili; quindi
  non e' correlato con il tipo di forget esistenziale usato qui

- forget usato per fare abduzione

  in logica proposizionale, primo ordine, descrittive e logic programming:

  . negare le manifestazioni M
  . aggiungerle alla formula
  . dimenticare tutto tranne le ipotesi possibili
  . negare il risultato

  funziona perche' FuE|=M e' lo stesso di Fu-M|=-E; proiettando Fu-M sulle
  ipotesi, si ottiene una formula che implica esattamente tutte le spiegazioni

  in tutti questi articoli, si parte sempre da una specifica manifestazione e
  si usa forget per ottenere le spiegazioni in questo modo; quindi forget e'
  uno dei passi per ottenere la spiegazione di una specifica manifestazione

  riassunto degli articoli, file e commenti espansi in local.bib

  delp-schm-19
  	e' il sistema detto sopra applicato a description logic; cita altri:

  	"[forgetting] has been proposed as a method for abduction in different
  	contexts (Doherty, Lukaszewicz, and Szalas 2001; Gabbay, Schmidt, and
	Szalas 2008; Wernhard 2013; Koopmann and Schmidt 2015b)."

  delp-schm-19-b
  	stesso metodo, applicato a una logica diversa (ALC+nominals)

  koop-schm-15
  koop-20
	sistema di forget in varie logiche descrittive; e' basato
	sull'osservazione che OuH|=Obs e' equivalente a Ou-Obs|=-H, dove O e'
	la teoria, Obs le osservazioni e H le sue spiegazioni; sulla base di
	questa osservazione, si ottiene H come il negato delle conseguenze di
	Ou-Obs

  wern-13
	abduzione in programmazione logica; il problema e' sempre quello della
	spiegazione (minima) di una specifica manifestazione; si parte dal
	sistema per la logica proposizionale di negare le manifestazioni,
	unire alla formula, proiettare e negare, dove la proiezione e' il
	contrario di forget; il risultato e' definito gwsc(); nel caso di
	programmi logici, dimostra che la definizione di spiegazione minima si
	riconduce attraverso delle trasformazioni che producono formule e
	insiemi di letterali proposizionali a una condizione basata su gwsc()

  Gabbay, Schmidt, and Szalas 2008
	citato da delp-schm-19, ma non trovato

	da articoli degli stessi autori sembra essere sempre su come eliminare
	quantificatori del secondo ordine, che e' il sistema usato anche
	nell'articolo di Doherty, Lukaszewicz, and Szalas 2001; abduzione e'
	solo una applicazione, e non viene nemmeno menzionata; non si parla di
	forget

  dohe-etal-01
	trovano le weakest sufficient conditions e le strongest necessary
	condition di una formula del primo ordine introducendo quantificatori
	del secondo ordine che poi eliminano; abduzione e' solo una
	applicazione di questo sistema; forgetting e' una applicazione di uno
	dei teoremi; non c'e' bisogno di citare

  lin-01
	vari risultati su wsc, weakest sufficient conditions; abduzione e' una
	loro applicazione; viene dato un metodo basato su implicati primi;
	dimostra che wsc si puo' ridurre a forget: si sostituisce la
	conclusione (manifestazione) con false nella formula, si dimenticano le
	altre variabili, si nega il risultato

- relazione di inferenza definita da abduzione: lobo-uzca-97

  l'articolo studia una relazione A|-B definita da un problema di abduzione, ma
  non e' la stessa cosa di A=>B; nel caso piu' semplice, A|-B e' vero se B e'
  implicato dalla disgiunzione di tutte le spiegazioni di A aggiunto alla
  teoria; in generale, c'e' una funzione F() che dato A produce una formula che
  lo spiega, e di questa formula si vede se implica B con la teoria

  questa relazione ha certe proprieta' nel caso piu' semplice e nel caso
  generale con la funzione F(); certe proprieta' valgono quando per abduzione
  si usa |=, e certe quando si usa una relazione di conseguenza |~ ma a seconda
  delle proprieta' di questa

  almeno nel caso piu' semplice |- si puo' esprimere come un modello
  cumulativo: (S,l) dove S e' un insieme di elementi detti stati e l() una
  funzione che assegna a ognuno un insieme di interpretazioni; da (S,l) e'
  possibile appunto definire una relazione |-

  non e' la stessa cosa perche' nell'articolo presente si cerca di esprimere =>
  invece di |-, ma e' correlato; si parla sempre di abduzione, si parla di una
  relazione binaria (|- o =>) che poi si puo' a seconda dei casi esprimere in
  un certo modo (con un modello comulativo o con un altro problema di
  abduzione); ma essendo su una diversa relazione binaria e' diverso del tutto

- espressione di abduzione con preferenze, pino-uzca-03

  viene definita una relazione |> che e' uguale => a parte l'inversione degli
  argomenti, ma che in piu' ha il vincolo che il secondo argomento deve essere
  una spiegazione del primo secondo la formula data (consistente e implica)

  a partire da quali proprieta' soddisfa |>, viene detto quali proprieta'
  soddisfa la corrispondente relazione di implicazione |~, ma questo non e'
  correlato; invece e' correlato il fatto che viene detto quali proprieta' deve
  avere |> perche' sia definibile a partire da un ordinamento <, e quali
  proprieta' di |> corrispondono a quali proprieta' di < e viceversa

  quindi si tratta sempre di esprimere => in altra forma, ma l'altra forma in
  questo caso e' un ordinamento e non un problema di abduzione; e le proprieta'
  non corrispondono

  - il vincolo che a|>b puo' valere solo se b e' una spiegazione abduttiva di a
    implica che e' possibile codificare in questo modo le spiegazioni abduttive
    e quelle minime per contenimento o cardinalita', ma non quelle del forget
    dato che e' dimostrato qui che queste non sempre sono spiegazioni del
    forget

  - non viene mai detto quali proprieta' |> dovrebbe avere perche' sia
    l'insieme delle spiegazioni di una formula, anche perche' lo sono sempre;
    viene dimostrato quando |> e' definibile da un qualche ordinamento, ma
    questo fa corrispondere |> a < mentre qui si vede quando => corrisponde a
    una formula con un ordinamento fissato (<= o c=)

  - pero' l'articolo e' correlato perche' |> viene rappresentato in altro modo,
    nello specifico come < o |~; pero' cambia l'altro modo, che in questo
    articolo e' sempre una teoria (classica o default) di cui viene fatto
    abduzione o abduzione minima secondo un ordinamento fissato

  l'articolo aigu-etal-18 e' una generalizzazione a logiche arbitrarie; non
  fornisce proprieta' che dovrebbe avere una explanatory relation per essere
  ottenibile a partire da altro; quindi non e' correlato a questo articolo

  l'articolo precedente pino-uzca-99 e' sulla corrispondenza fra explanatory
  relation e relazione di inferenza; l'unico risultato correlato e' che viene
  detto quali proprieta' deve avere una explanatory relation perche' sia
  esprimibile in un certo modo a partire da una relazione di inferenza

- framework generale di forget, beie-etal-19

  definiscono un framework astratto su cui fare abduzione ma non formalizzano  
  le operazioni di forget su di esso; sembrano usarlo solo per gli esempi, poi
  passano direttamente agli ocf

  esiste una nozione astratta di stato doxastico, ma cosa ci si fa non viene
  detto tranne che nel caso specifico degli OCF; vengono definite varie forme
  di forget; quelle che piu' si adattano ad abductive forgetting sembrano
  essere astrazione, marginalizzazione e focusing

  abstraction:

    e' una forma che sembra assomigliare a forget; se P e' lo stato doxastico,
    allora:

  	P |= A e B implicano C
	P |= A e -B implicano C
    producono per astrazione:
	P |= A implica C

    in abduction l'implicazione e' al contrario, quindi sarebbe C=>AB e C=>A-B,  
    che in abductive forgetting non produce C=>A, ma non richiede che ci siano
    tutti e due
  
  marginalization:

    e' piu' simile a forget, perche' si focalizza su una parte del linguaggio
  
    dato pero' che non hanno detto cosa si fa nel caso astratto ma solo
    nell'esempio concreto, non e' possibile fare un confronto

  focusing:

    e' la forma di marginalizazzione in cui viene determinato automaticamente
    l'alfabeto su cui restringere il linguaggio
  
    questo non e' affatto quello che si fa in abductive forgetting, pero'
    avrebbe anche senso considerarlo; nell'articolo e' citato come caso
    interessante per studi futuri (conclusions.tex)

articoli rilevanti
------------------

- altri tipi di abduzione o minimalita', paul-93 eite-gott-95-a

  si possono citare se richiesto di menzionare l'esistenza di tipi di abduzione
  che non sono logic-based e altri tipi di minimalita'

- abduzione su una teoria di default, eite-etal-97, tomp-03

  definizione di abduzione uguale alla definition 2

  una applicazione di abduction from default e' planning; anche per planning ha
  senso una definizione di forget?

- forgetting in answer set programming - a survey, gonc-etal-21

  citazione a pag 2 beierle per motivazione aggiuntiva forget; altre simili
  sempre a pagina 2; applicazioni di forgetting a pagina 35

  nel forgetting in logic programming is mantengono gli answer set o gli
  ht-models piu' che direttamente le conseguenze; ma si tratta comunque di
  modelli, usati per trarre conclusioni

- abduzione su description logics, else-etal-06

  attualmente non citato; si puo' citare nel contesto di abduction applicato ad
  altre logiche; per il resto non c'e' molto altro di collegato a forget; una
  differenza significativa e' che la simple concept abduction e' definita da G
  |= H c= C, ma e' essenzialmente sempre un modo di dire che H implica C
  insieme a G

  riassunto delle regole di abduzione (consistency, minimality, relevance,
  explanatoriness)

\vv

\enddraft

\def\c#1{\accent 24#1}
\newcommand{\etalchar}[1]{$^{#1}$}

\bibliographystyle{alpha}

\end{document}